\documentclass[aps,pra,a4,twocolumn]{revtex4-1}
\usepackage{amsmath}
\usepackage{amsfonts}
\usepackage{amssymb}
\usepackage{graphicx}
\usepackage{color}

\begin{document}

\title{Fast trimers in one-dimensional extended Fermi-Hubbard model}
\author{A. Dhar$^1$, P. T\"orm\"a$^2$ and J. J. Kinnunen$^2$}
\affiliation{$^1$ Institut  f\"ur  Theoretische  Physik,  Goethe-Universit\"at  Frankfurt  am  Main, Max-von-Laue-Stra\ss e  1,  60438  Frankfurt  am  Main,  Germany\\
$^2$ COMP Centre of Excellence and Department of Applied Physics, Aalto University, FI-00076 Aalto, Finland}

\begin{abstract}
{We consider a one-dimensional two component extended Fermi-Hubbard model with nearest neighbor interactions and mass imbalance between the two species. We study the stability of trimers, various observables for detecting them, and expansion dynamics. We generalize the definition of the trimer gap to include the formation of different types of clusters originating from nearest neighbor interactions. Expansion dynamics reveal rapidly propagating trimers, with speeds exceeding doublon propagation in strongly interacting regime. We present a simple model for understanding this unique feature of the movement of the trimers, and we discuss the potential for experimental realization.}
\end{abstract}

\maketitle

\section{Introduction}

Formation of clusters is at the heart of all matter around us. 
Interparticle interactions lead into formation of pairs, trimers, and clusters, but what exactly is required for larger clusters to form? 
The paradigmatic model in condensed matter physics is the Hubbard model, describing low-energy particles moving in a periodic potential and interacting through short-range interactions.
However, for fermionic particles and on-site interactions, the Fermi-Hubbard model has been shown to support trimers only in a limited range of parameters~\cite{takahashi_excitonic_1970}. 
Long-range interactions are required for stabilizing larger clusters.
Indeed, extended Fermi-Hubbard model involving nearest-neighbor interactions is the next simplest model, but one that is sufficiently rich to describe the formation of large clusters over a large range of parameters.
%So what kind of clusters does the model predict? And what are their properties?

In the last two decades, there has been significant progress in cooling and trapping ultracold atoms in different lattice geometries leading to the simulation of a variety of quantum models~\cite{Bloch_review2017} including the extended Bose-Hubbard model~\cite{Baier_extended_2016}. 
Indeed, in addition to short-range on-site interactions, it is now possible to engineer also long-range interactions using ultracold molecules, Rydberg-dressed atoms and dipolar atoms~\cite{Lahaye_the_physics_2009,Park_molecules,Yan_molecules,Schauss_Rydberg,Baier_extended_2016,Ferlaino_rotons}. 
Trapping dipolar atoms in optical lattices to form itinerant models have recently been achieved experimentally~\cite{Baier_extended_2016}. 

Another important parameter for cluster formation is the mass imbalance between the particles.
Presence of mass imbalance breaks the SU(2) symmetry in the system explicitly giving rise to a variety of novel quantum phases such as unconventional superconductivity, exotic forms of quantum magnetism and trimer phases.~\cite{Casalbuoni_RMP,Auerbach_book,Orso_luttinger_2010,Orso_fermionic_2011}
There are several ways to induce mass imbalance in optical lattices such as differential coupling of the laser field to different atom transitions~\cite{Lamporesi_mass_imbalance_experiment-1}, applying an oscillatory magnetic field gradient~\cite{Jotzu_creating_2015}, or using different atomic species or isotopes~\cite{Ferlaino_ErDy}. Effects of mass imbalance in continuous systems have been extensively studied both theoretically and experimentally, revealing the existence of Efimov trimers and Kartavtsev-Malykh trimers at different regimes of mass ratios and interactions~\cite{Efimov_1973,Malykh_efimov,Ueda_2012,Grimm_2014,Pascal_2016}.

We study trimer formation and dynamics in two-component fermionic dipolar gas in a one-dimensional optical lattice.
Using the extended Hubbard model, in which the fermionic atoms interact through nearest-neighbor interactions, our aim is to understand cluster formation in a few-body system, the stability of the trimer configuration, and how the trimers move in the lattice. We perform both static and dynamic analysis.

In the static analysis, we consider relatively weak interactions, resulting in spatially very large trimers.
Depending on the mass-ratio of the two atomic components~\cite{Cazalilla_two-component_2005,Wang_quantum_2009,Burovski_multiparticle_2009,Chung_multiple_2017,Sekania_mass-imbalanced_2017}, these trimers can be understood to consist either of a single light atom binding two heavier atoms together by mediating an effective attractive long-range interaction between the heavy atoms~\cite{Mathy_trimers_2011, Jag_observation_2014}, or, in the opposite mass-ratio limit, a single heavy atom providing a pinning potential for two light atoms.
We provide a general method for determining the trimer gap, and we consider also the use of two-particle correlators for determining the stability of trimers.

In the dynamic studies, we consider more strongly bound trimers, and analyze in detail the movement of an initially localized trimer and determine how the propagation speed depends on hopping parameters and the strength of the nearest-neighbor interaction.
We show that, despite the inherent complexity of transporting a multiparticle object in the lattice, the trimers can actually propagate faster than corresponding dimers. 
We provide simple description of both dimer and trimer propagation that yields excellent agreement with the essentially exact numerical results provided by matrix product states (MPS) method.

The results are immediately relevant for experiments on long-range interacting ultracold atoms in optical lattices.
We consider several one- and two-particle correlators for analyzing the presence and propagation of trimers, and we discuss the possibility of using these in actual experimental settings.

The structure of the work is the following: in Section~\ref{sec:hamiltonian} we describe the theoretical model used for studying the system.
In Section~\ref{sec:static} we describe the static properties of the trimers: trimer gap and the size of the trimer. These are studied in weakly interacting systems, since that is the regime where the trimer stability is nontrivial, as strong attractive long-range interaction tends to lead into formation of large clusters. In contrast, in Section~\ref{sec:trimer_dynamics}
we study the propagation of trimers and dimers in the lattice by releasing initially tightly trapped atoms into a larger homogeneous lattice. These studies are done with strong interactions to have more stable trimers that can survive the release without breaking.
We show the surprising result that strongly bound trimers can move faster than dimers along the one-dimensional lattice, and we explain the finding with a simple model that involves near-degenerate trimer states. 
Section~\ref{sec:experiments} is devoted for analyzing the experimental relevance of the results.
The intriguing results regarding rapidly moving trimers should be within easy reach experimentally as there is no need for low temperatures.  
We conclude by summarizing the key results in Section~\ref{sec:conclusion}.

\section{Model Hamiltonian}
\label{sec:hamiltonian}

We consider a two-component fermionic system in a lattice with nearest-neighbor interactions which can be described by the extended Fermi Hubbard model. Additionally, we consider the two fermionic species to possess different masses, reflected in the different tunneling amplitudes. The Hamiltonian describing such system is
\begin{eqnarray}
H&=&-\sum_{\langle ij \rangle, \sigma}{t_{\sigma}(c_{i,\sigma}^{\dagger}c_{j,\sigma}+\mathrm{h.c.})}+U\sum_in_{i, \uparrow}n_{i, \downarrow} \\ \nonumber &&+\sum_{\langle i,j \rangle, \sigma}V n_{i,\sigma}n_{j,\sigma}+\sum_{\langle i,j \rangle} V' n_{i, \uparrow}n_{j, \downarrow}.
\end{eqnarray}
Here, $c_{i,\sigma}(c_{i,\sigma}^{\dagger})$ are the annihilation (creation) operators at site $i$ of fermionic species $\sigma(=\uparrow/\downarrow)$. The first term denotes the tunneling to the nearest neighbor sites with a spin dependent tunneling amplitude, $t_{\sigma}$. The second term expresses the on-site interaction between opposite spins. The third and fourth term takes into account the nearest neighbor interactions between the same and different species respectively. 

We concentrate on the experimentally more relevant case in which inter- and intraspecies nearest-neighbor interactions are equal, i.e. $V=V'$.
In addition, the on-site interaction $U$ is assumed to be small, although we also discuss the effect of relaxing both of the assumptions.

The hopping ratio is defined as $t_\mathrm{ratio}=t_{\downarrow}/t_{\uparrow}$, where we have fixed $t_{\uparrow}=1$, and varied $t_{\downarrow}$.
That means that the mass of $\downarrow$-atoms is changed from heavy (for small $t_\mathrm{ratio}$) to light (for $t_\mathrm{ratio}$ close to or above unity).
For static simulations, the lattice size has been kept to 100 sites, whereas for the dynamics studies
the typical lattice size is 53 sites.

To study the above mentioned system in one dimension, we have used the matrix product states (MPS) method which has proved to be extremely accurate in lower dimensions for both static and dynamic investigations~\cite{MPS_NJP, MPS_arxiv,MPS_sourceforge}, we have used the maximum bond dimensions to be 1000 resulting in an error of less than $10^{-12}$. For dynamical studies, the Krylov-based time evolution~\cite{saad_krylov1992} was used with a tolerance in the Lanczos procedure to determine the matrix exponential set to less than $10^{-6}$ and maximum bond dimensions to 500.   

\section{Static trimer: energy and correlations}
\label{sec:static}

For studying individual trimers, we consider a single $\uparrow$-atom and two $\downarrow$-atoms.
We concentrate on the case of a light $\uparrow$-atom, which then effectively mediates an interaction between the two $\downarrow$-atoms. 
Such a setting has been studied earlier assuming only on-site interactions. For the case of equal masses, or the hopping ratio $t_\mathrm{ratio}=1$, the system with only on-site interactions has been shown to have no trimers~\cite{takahashi_excitonic_1970}. However, for mass-imbalanced systems, the trimer formation has been studied and predicted~\cite{Orso_luttinger_2010,Roux_multimer_2011, Dalmonta_dimer_2012} to occur over a large range of parameters.

Earlier studies in one-dimensional extended Hubbard model using the quantum Monte Carlo method found the tendency of atoms forming large clusters~\cite{Batrouni_pair_2009}.
Since in this work we consider systems of at most three atoms, the formation of larger clusters is not included. 
However, even if the ground state of the system was one big cluster, in practice the system would consist of multiple smaller clusters, as the binding energies decrease with increasing cluster size in one-dimensional lattice. Trimers are therefore highly relevant, even if the true ground state would be something quite different. In such case, it will be very interesting to see how the various cluster configurations propagate. Our results suggest that, with suitable interaction parameters, the trimers and singlons (i.e. single particles) are the only fast moving configurations.

%Trimer formation due to long-range dipole-dipole interactions have been studied also in bilayer systems~\cite{Zinner_few-body_2011}
%with long-range interaction

\subsection{Trimer energy}

An important quantity that characterizes the behavior of trimers is the trimer gap, defined as the energy needed to break a single trimer. 
The trimer gap is usually defined as~\cite{Orso_luttinger_2010,Dalmonta_dimer_2012,Roux_multimer_2011}
\begin{eqnarray}
\label{eqn:trimer-gap}
\Delta_\mathrm{tr}&=&-\lim_{L\rightarrow\infty}[E_L(N_{\uparrow}+1, N_{\downarrow}+2)+E_L(N_{\uparrow}, N_{\downarrow}) \\ \nonumber &&-E_L(N_{\uparrow}+1, N_{\downarrow}+1)-E_L(N_{\uparrow}, N_{\downarrow}+1)]
\end{eqnarray}
where $E_L(N_{\uparrow}, N_{\downarrow})$ is the ground state energy with spin populations $N_{\uparrow}, N_{\downarrow}$ in a system of $L$ sites. 
It is essential to explain the above definition of trimer gap before we proceed to describe how it has to be modified in the presence of nearest neighbor interactions. 
The first term on the right hand side in Eq.~\eqref{eqn:trimer-gap} denotes the energy of a system of $N_{\uparrow}, N_{\downarrow}$  atoms and a trimer consisting of one $\uparrow$ and two $\downarrow$ atoms. 
This trimer can break into a dimer consisting of one $\uparrow$ and one $\downarrow$ atoms, and a singlon of one $\downarrow$ atom. 
The third and fourth terms on the right hand side denote the energies of $N_{\uparrow}, N_{\downarrow}$ atoms and a doublon, and $N_{\uparrow}, N_{\downarrow}$ atoms and a singlon respectively. 
The trimer gap of a single trimer will thus be obtained by adding the energy for $N_{\uparrow}, N_{\downarrow}$ as given by the second term. 
Unless one is interested in many-body effects, one can choose $N_\uparrow = N_\downarrow = 0$.
%The above definition works well only if one can assume that the trimer can break into doublons consisting of 
%opposite spins only. 
However, the definition in Eq.~\eqref{eqn:trimer-gap} misses the possibility of formation of larger clusters, but also the many different possibilities of the trimer to break into. 
This is particularly important in the presence of nearest-neighbor interactions, in which case the trimer can break into a dimer consisting of two $\downarrow$ atoms situated at neighboring sites and a singlon of $\uparrow$ atom.
Furthermore, nearest-neighbour interactions make large cluster formation much more likely, and hence there is need for generalizing the trimer gap to larger clusters.

Here we define the trimer gap as the lowest energy separation of any possible combination which conserves the particle numbers.
That is, we solve the energy spectrum of the ground state energies of all combinations $E_L(N_1,N_2)$ for all $N_1 \in \left[0,\ldots,N_\uparrow\right]$ and $N_2 \in \left[0,\ldots,N_\downarrow\right]$.
These energies are then summed pairwise such that the sum of the particle numbers of the two configurations equals $N_\uparrow$, $N_\downarrow$:
\begin{equation}
\label{eq:ground_state_spectrum}
   E_L(N_\uparrow-N_1,N_\downarrow-N_2) + E_L(N_1,N_2).
\end{equation}
In the case of trimers studied in this work, i.e. for $N_\uparrow = 1,N_\downarrow = 2$, this corresponds to the total energy of a configuration in which the trimer is broken into two parts, one with $N_1$ $\uparrow$-atoms and $N_2$ $\downarrow$-atoms, and the other part containing the rest of the atoms.
By calculating these for all combinations $N_1 = 0,1$ and $N_2 = 0,1,2$, one can identify the trimer gap as the energy separation between two of the lowest energies.
%Assuming that the lattice is sufficiently large, the lowest energy state is necessarily the trimer state, or the configuration with $N_1=N_2=0$.
Notice that this definition works also for larger clusters, giving the energy spectrum of all partitions of the system into two parts.

\begin{figure}[h!]
  \centering
  \includegraphics[width=0.47\linewidth]{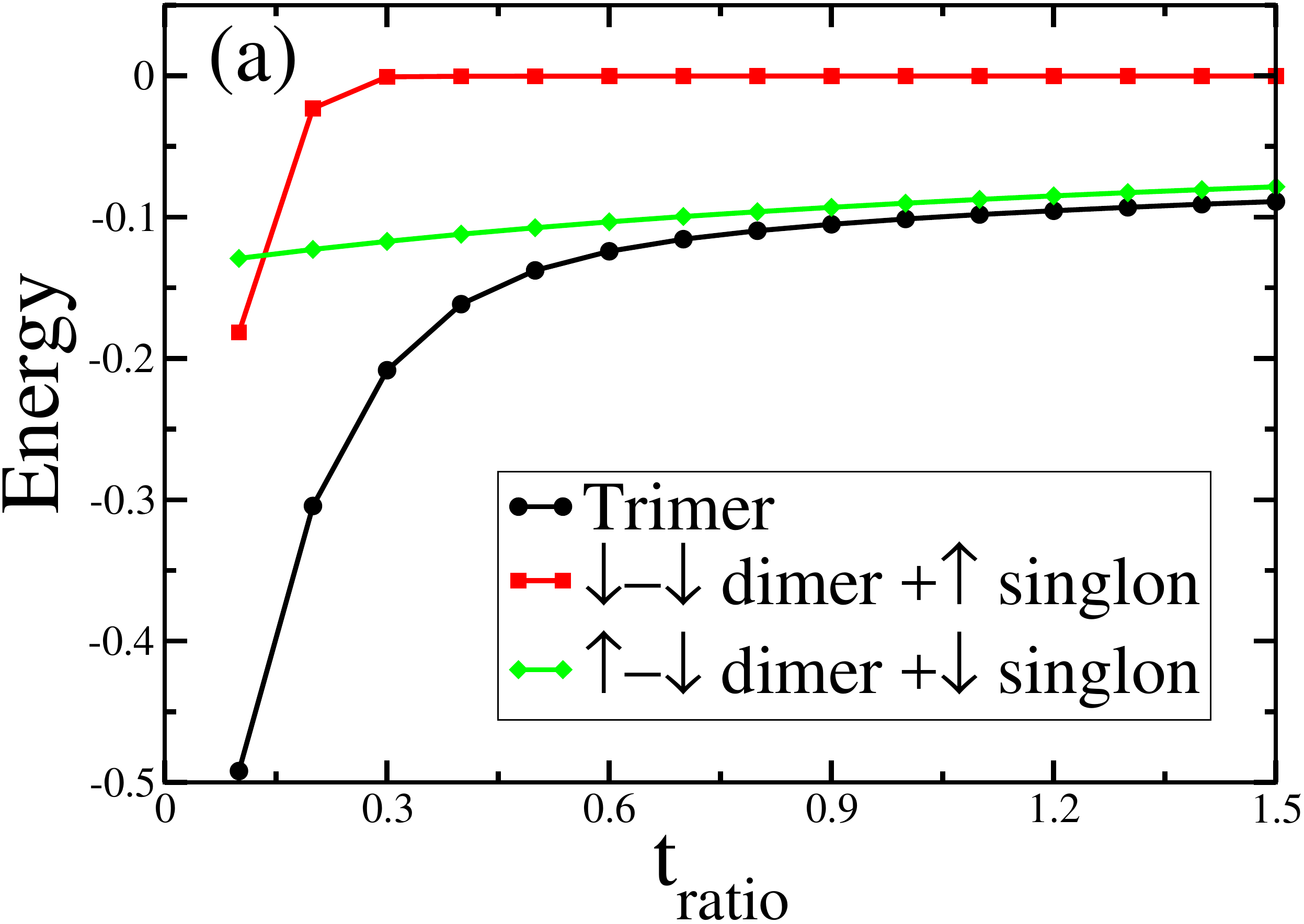}
  \includegraphics[width=0.47\linewidth]{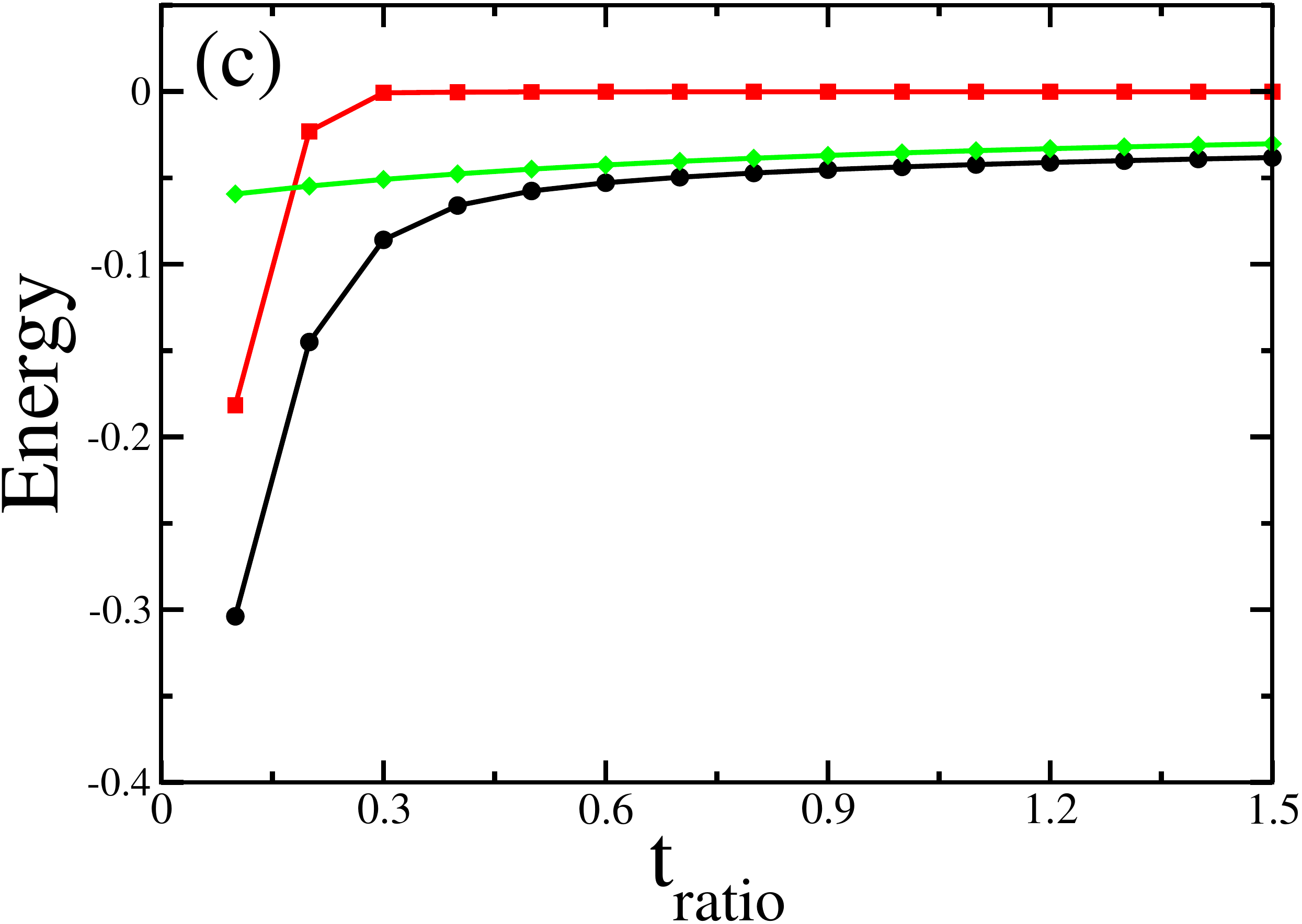}
  \includegraphics[width=0.47\linewidth]{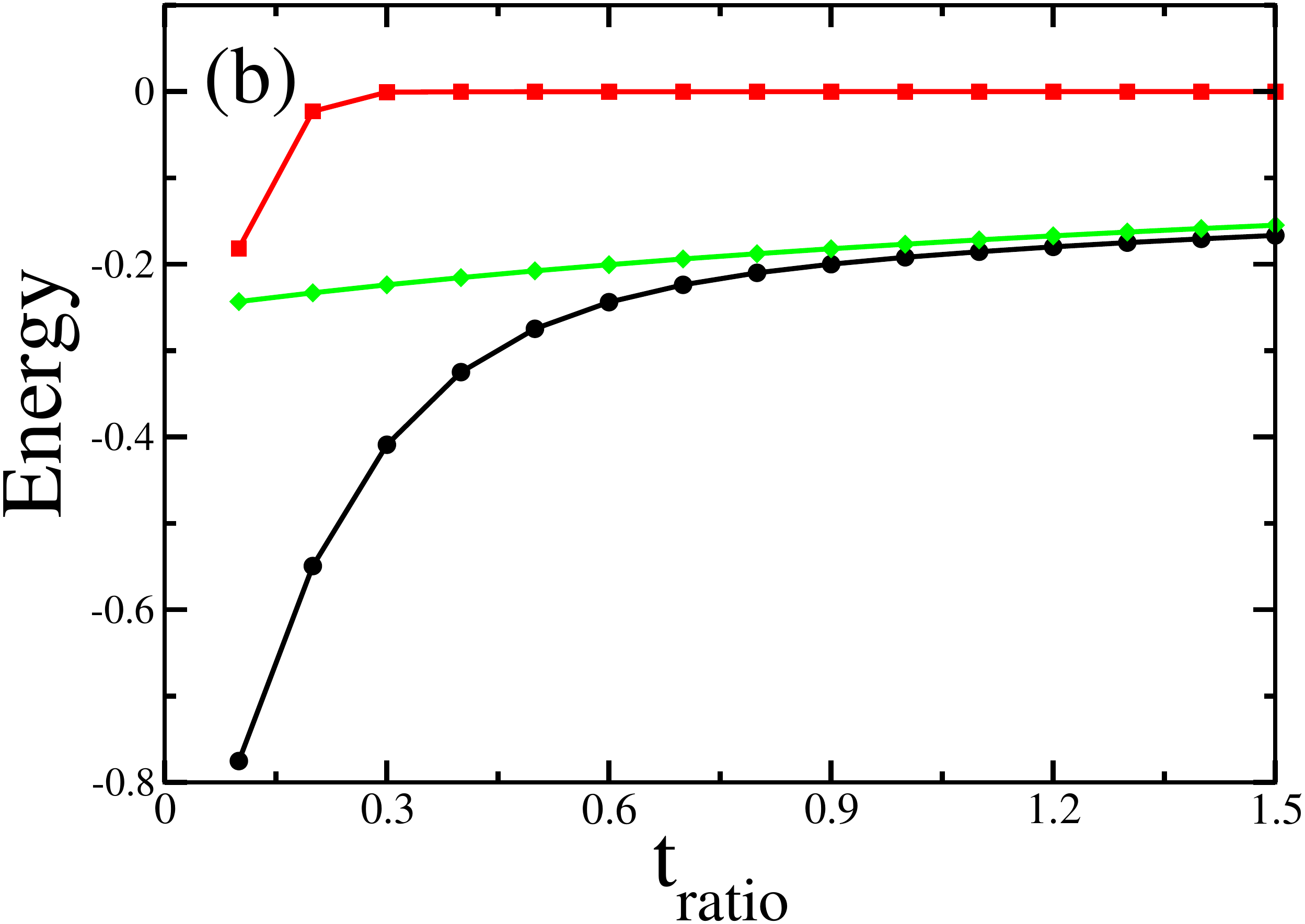}
  \includegraphics[width=0.47\linewidth]{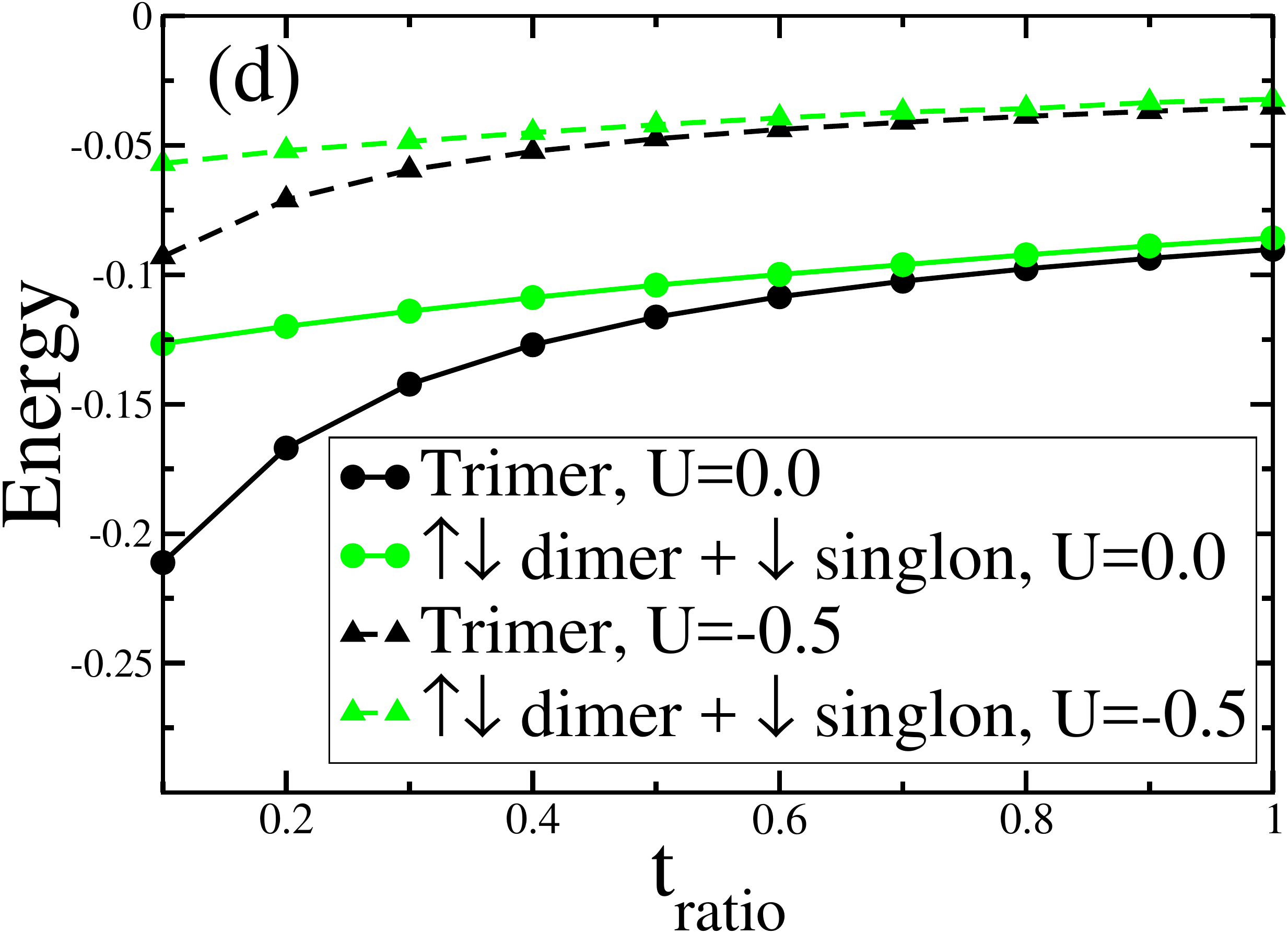}
  \caption{Energy of a single trimer. (a) $V=V'=-0.5\,t_\uparrow, U=0$, (b) $V=V'=-0.5\,t_\uparrow, U=-0.5\,t_\uparrow$, (c) $V=V'=-0.5\,t_\uparrow, U=0.5\,t_\uparrow$, having the same legends, and (d) $V=-0.5\,t_\uparrow, V' = 0$, $U$. Shown are the energies of a trimer (configuration $N_\uparrow = 1, N_\downarrow=2$), an $\uparrow-\downarrow$ dimer separated from a single atom (configurations $N_\uparrow = 1, N_\downarrow=1$ and $N_\uparrow = 0,N_\downarrow = 1$), and a $\downarrow \downarrow$ dimer separated from a single atom (configurations $N_\uparrow = 0, N_\downarrow=2$ and $N_\uparrow = 1,N_\downarrow = 0$), except for the last case in which the intraspecies dimer is unstable in absence of interactions. The separation of the two lowest lines is the trimer gap. In order to better show the effect of interactions, the total energy of the corresponding non-interacting system (roughly equal to $-t_\uparrow - 2t_\downarrow$) has been subtracted from all energies.}
  \label{fig:one_trimer_energy}
\end{figure}

Fig.~\ref{fig:one_trimer_energy} shows the energy of a single trimer as a function of the hopping ratio $t_\mathrm{ratio} = t_\downarrow/t_\uparrow$ for different combinations of on-site $U$ and nearest-neighbour interactions $V,V'$.
The trimer gap (energy gap between the trimer and lowest-lying dimer configuration) is large for low hopping ratios, where the kinetic energy of the two $\downarrow$-atoms is small. However, with increased mobility, the trimer becomes less strongly bound, shown as decreasing trimer gap.
Fig.~\ref{fig:one_trimer_energy} shows also the dependence of the trimer and dimer energies on the on-site interaction strength $U$. 
As one would expect, repulsive on-site interaction $U$ makes trimers and interspecies dimers less strongly bound.
However, qualitatively the ground state properties of the trimer do not seem to depend on the choice of on-site interaction $U$. 

The trimer configuration has the lowest energy for all cases we have studied. 
However, since the energy of the doublon+singlon configuration is done by calculating the energies of the doublon and singlon independently, the density-dependent Hartree energy is ignored.
Hartree energy should be present even without actual pair formation and it can be estimated for a $\downarrow$-singlon interacting with an $\uparrow \downarrow$-dimer as
\begin{equation}
  \Delta \Sigma_\mathrm{Hartree} = \frac{2V+2V'}{L},
\end{equation}
where the interaction terms $2V$ and $2V'$ arise from interactions of the singlon with both of the atoms comprising the dimer and the prefactor 2 from the nearest-neighbour range of the interaction. The factor $1/L$ comes from the mean-field densities, assuming even distribution over a lattice of $L$ sites. As an example, for $V=V'=-0.5\,t_\uparrow$ and $L=53$ we have $\Delta \Sigma_\mathrm{Hartree} = -2.0\,t_\uparrow/53 \approx 0.038\,t_\uparrow$.
In the partitioning scheme used here, the trimer can be considered stable if the calculated trimer gap is larger than the mean-field Hartree energy shift.
Hence, for the weak interactions considered in Fig.~\ref{fig:one_trimer_energy} the trimer state can be considered stable at least for $t_\uparrow/t_\downarrow < 0.4$. However, for larger hopping ratios the picture is less clear and there is need for other criteria besides energy spectrum for determining the stability of the trimer state.
We do expect lattice size to have some effect in the weakly bound trimer regime. 
Due to numerical complexity of the calculations, we have been able to do only preliminary finite size scaling studies and hence our calculations are not sufficient for determining actual phase diagram in this regime.

Finally, notice that the energy spectrum considered here involves only \emph{ground-state energies} of various configurations. One can, and will, have multiple excited trimer states. These will be important for the trimer dynamics, as will be discussed in section~\ref{sec:trimer_dynamics}.

\subsection{Trimer size}

Another possibility for identifying trimers is to consider various two-particle correlators. 
The asymptotic long-range behaviour of the doublon correlator $\langle c_{i,\downarrow}^\dagger c_{i,\uparrow}^\dagger c_{i+j,\uparrow} c_{i+j,\downarrow}\rangle$ has been shown to yield information of the trimer state in the case of on-site interactions~\cite{Orso_luttinger_2010}. For trimers the correlator decays exponentially whereas for dimer state only algebraically. However, the doublon correlator involving only atoms in the same site does not describe very well dimers of atoms interacting through nearest-neighbour interactions, as is the case in present work.

\begin{figure}[h]
  \centering
  \includegraphics[width=0.47\linewidth]{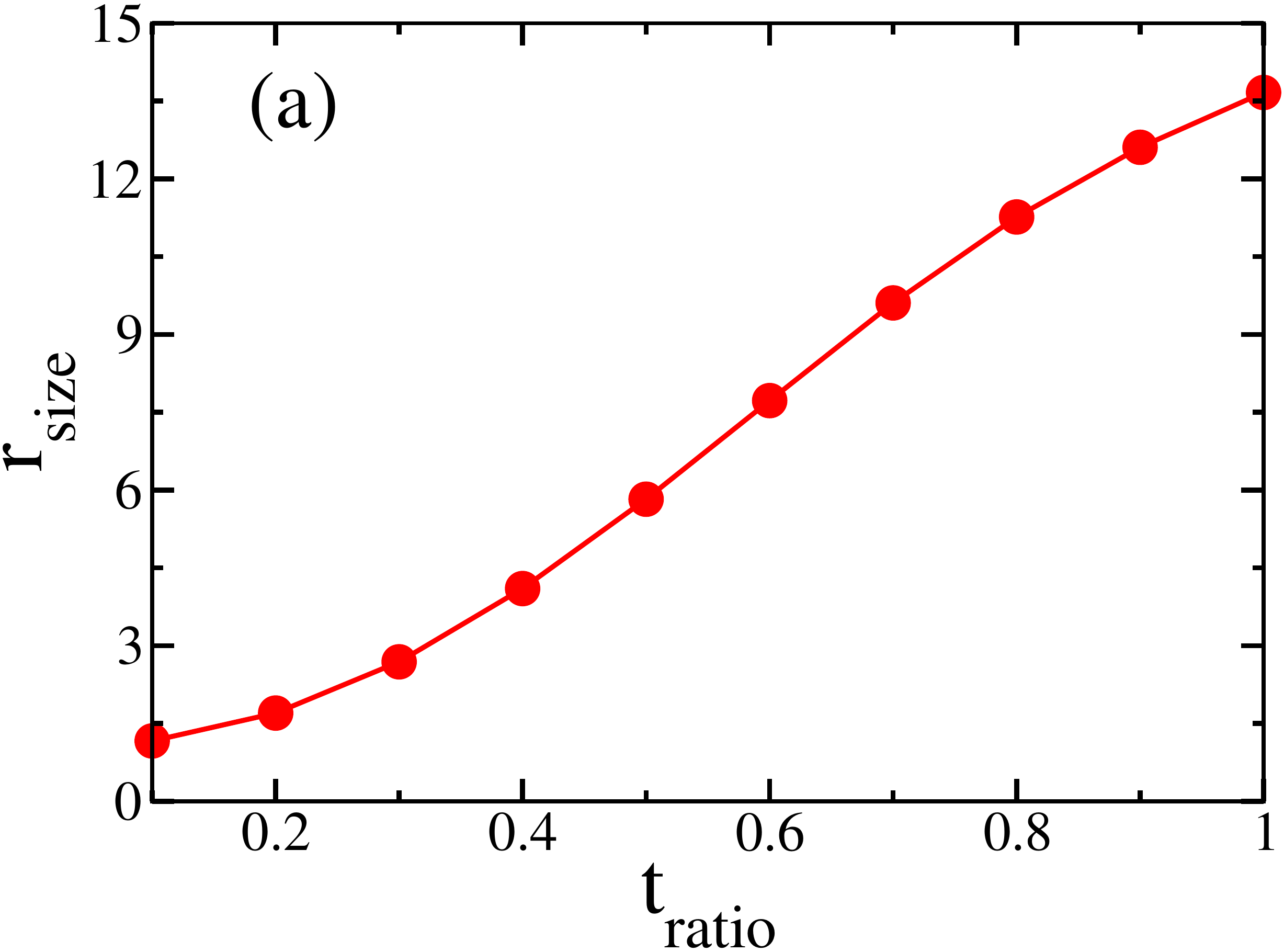}
  \includegraphics[width=0.47\linewidth]{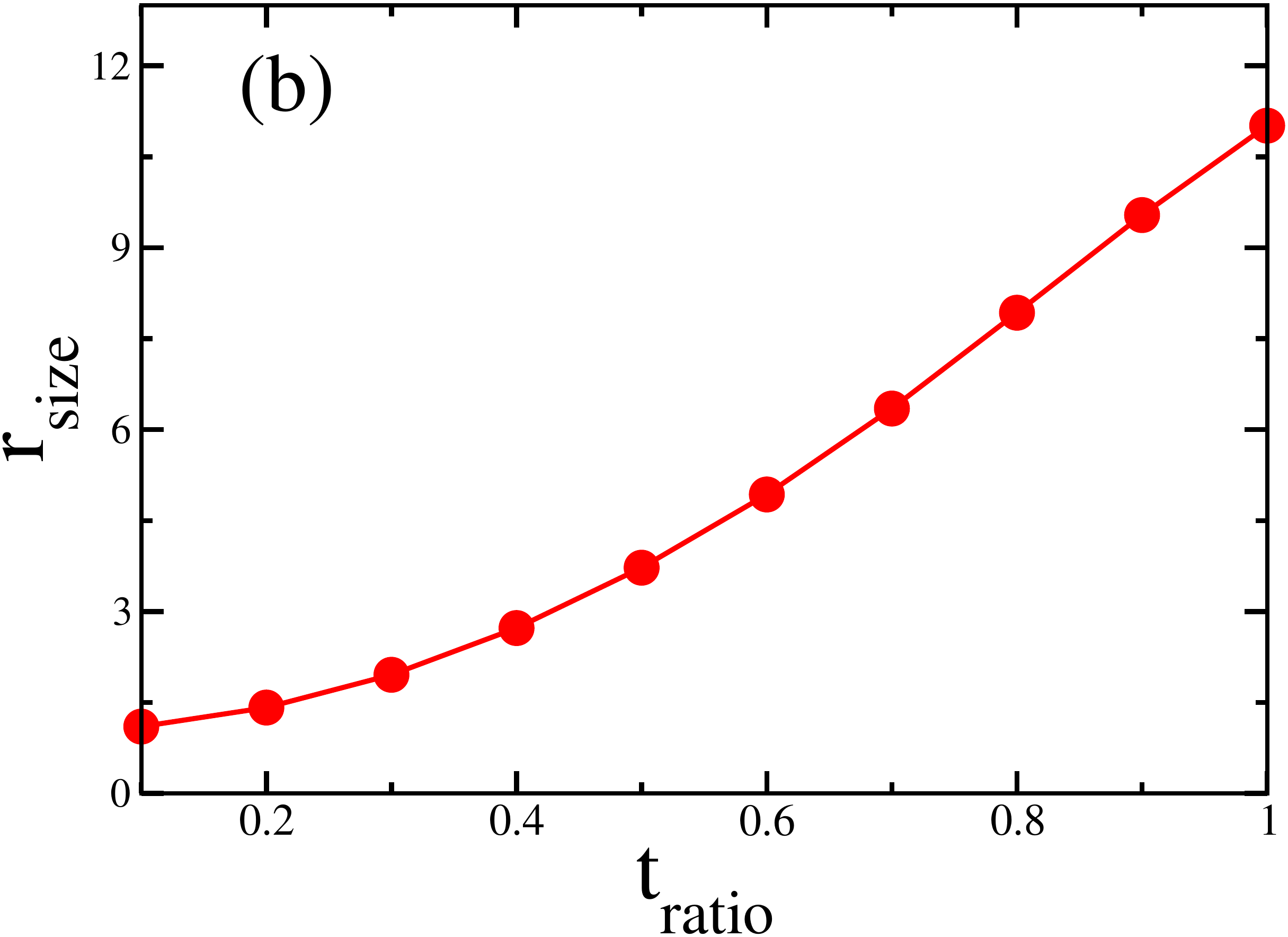}
  \includegraphics[width=0.47\linewidth]{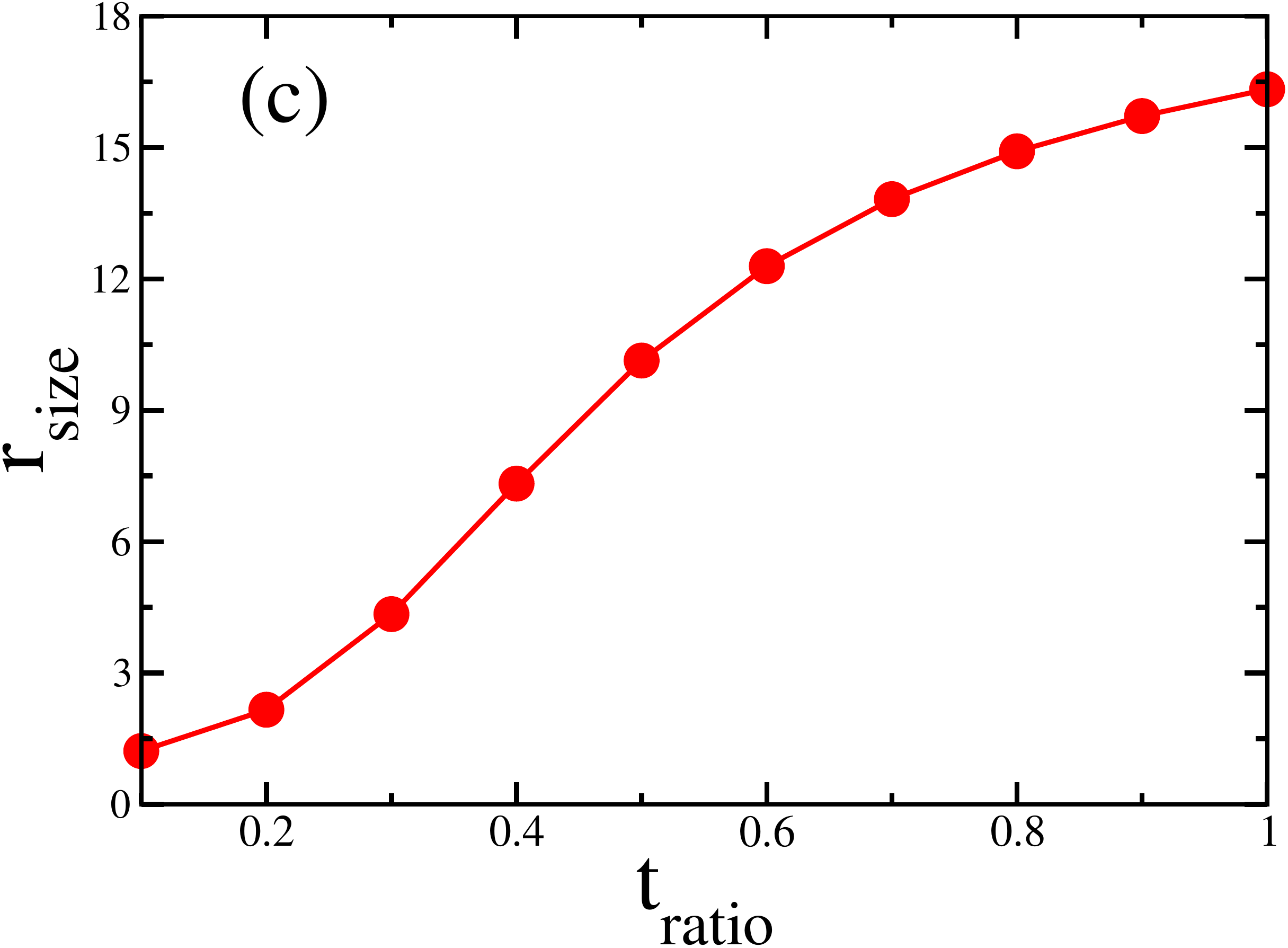}
  \includegraphics[width=0.47\linewidth]{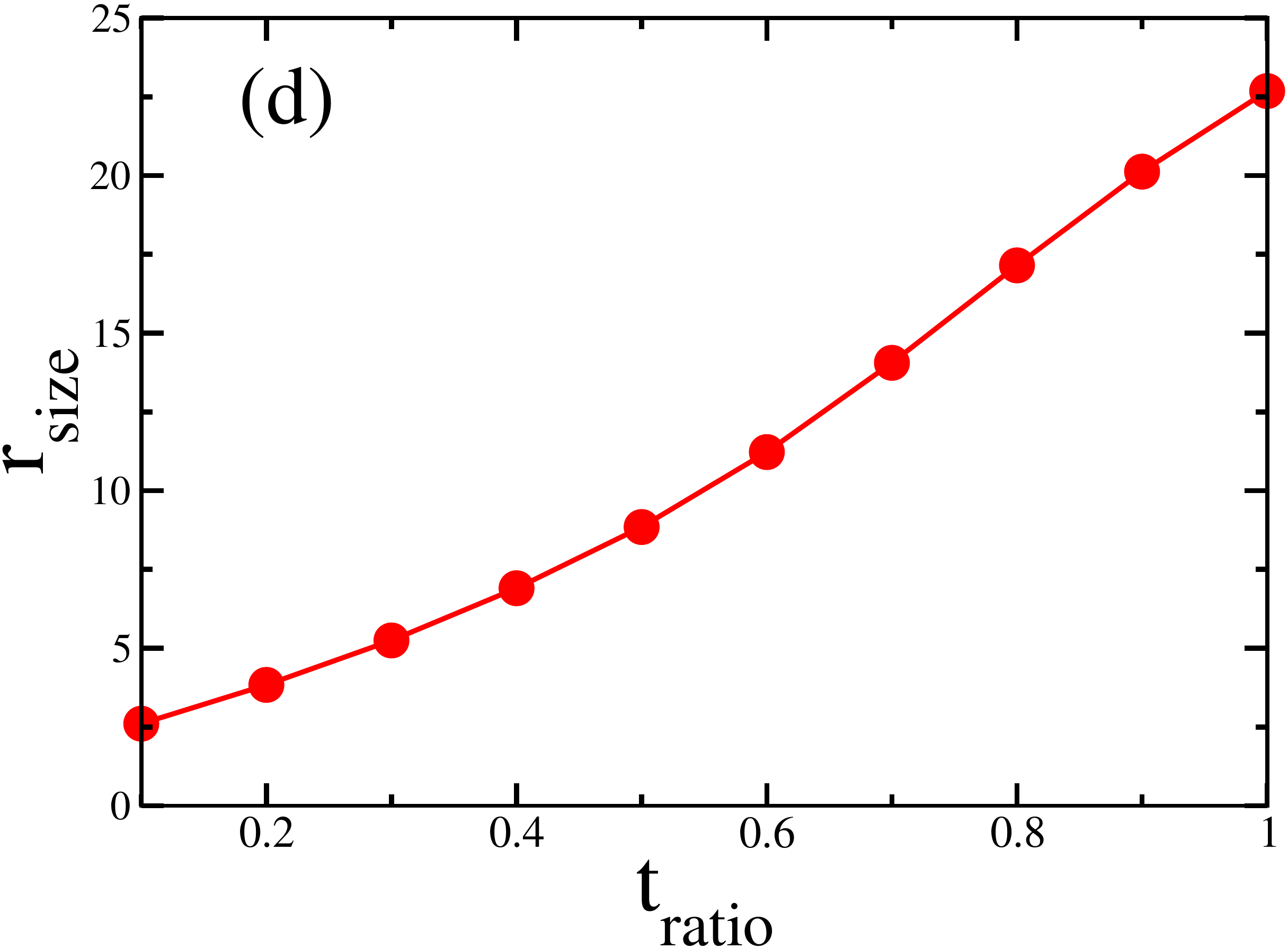}
  \caption{Average size of a trimer. (a): $V=V'=-0.5\,t_\uparrow, U=0$, (b): $V=V'=-0.5\,t_\uparrow, U=-0.5\,t_\uparrow$, (c): $V=V'=-0.5\,t_\uparrow, U=0.5\,t_\uparrow$,  and (d): $V=-0.5\,t_\uparrow, V' = 0$, $U=0$.}
  \label{fig:trimer_size}
\end{figure}

Instead of the on-site doublon correlator, we can try to identify trimers by considering the distance of the two $\downarrow$-atoms. The average distance is a measure of the size of the trimer.
That is, we define
\begin{eqnarray}
  r_\mathrm{size}=\frac{\sum_{i,d}{\langle n_{i, \downarrow} n_{i+d, \downarrow}\rangle \times d}}{\sum_{i,d\neq 0}{\langle n_{i, \downarrow} n_{i+d, \downarrow}\rangle }}
  \label{eq:trimer_size}
\end{eqnarray}
Fig.~\ref{fig:trimer_size} shows the calculated trimer size for a few chosen weakly interacting configurations.

Comparing with the calculated trimer energies in Fig.~\ref{fig:one_trimer_energy} we see that as the trimer becomes less strongly bound, the trimer size grows accordingly.
Due to weak interactions considered in these static calculations, the trimer size is surprisingly large, with the two majority atoms being even over 10 lattice sites away on average.
This is the case also in the absence of intraspecies interactions, see Fig.~\ref{fig:trimer_size} d).
In this case the interaction binding the two $\downarrow$ atoms has to be provided by the mediating $\uparrow$-atom. 

To obtain some scale for the average distance, one can consider a simple uncorrelated system, in which atoms are not interacting and quantum statistics is ignored. In such case, all correlators are equal $\langle n_{i,\downarrow} n_{i+d,\downarrow} \rangle = \langle n_{i,\downarrow} \rangle \langle n_{i+d,\downarrow} \rangle=L^{-2}$, and Eq.~\eqref{eq:trimer_size} can be solved analytically. For large $L$, one obtains $r_\mathrm{size}^\mathrm{uncorrelated} = L/3$, which for the lattice of $L=100$ sites yields average distance of $33$.
Clearly the average distances shown in Fig.~\ref{fig:trimer_size} are lower than this. 
Indeed, it appears that average distance is a better signature of bound trimers than the trimer gap.

Notice that the $\downarrow-\downarrow$-correlator yields also the size of the $\downarrow-\downarrow$-dimer, and is therefore not a very good way for distinguishing the trimer state from the $\downarrow-\downarrow$-dimer. However, as seen from the energy spectrum, the trimer state is mainly competing with the $\uparrow-\downarrow$-dimer state in the weakly bound regime.

In the following we will consider trimer dynamics with much more deeply bound trimers (originating from interactions of the order $V \sim -10\,t_\uparrow$).
%Using much stronger nearest-neighbour interactions, the trimer state is well-defined.
Even for hopping ratio close to unity, such strongly bound trimers have typical sizes of at most a few lattice sites. %, and hence we do not expect finite size effects to play a role there.

%\begin{figure}
%  \centering
%  \includegraphics[width=0.8\columnwidth]{../figs/plot-average-size-down-down-U0-V05-L100-one-trimer.pdf}
%  \caption{Size of one trimer for U=0, V=-0.5. Blue line is as earlier, for the red line we have now corrected the definition of the average size by removing the $d=0$ term from the denominator.}
%\end{figure}

%\begin{figure}
 % \centering
%  \includegraphics[width=0.32\columnwidth]{../figs/fig_avg_distance_U025_V025_L100.pdf}
%  \includegraphics[width=0.32\columnwidth]{../figs/fig_avg_distance_U025_V025_L150.pdf}
%    \includegraphics[width=0.32\columnwidth]{../figs/fig_avg_distance_U025_V025_L200.pdf}

%  \includegraphics[width=0.32\columnwidth]{../figs/fig_avg_distance_U0_V025_L100.pdf}
%  \includegraphics[width=0.32\columnwidth]{../figs/fig_avg_distance_U0_V025_L150.pdf}
%    \includegraphics[width=0.32\columnwidth]{../figs/fig_avg_distance_U0_V025_L200.pdf}

%  \includegraphics[width=0.32\columnwidth]{../figs/fig_avg_distance_U0_V05_L100.pdf}
%  \includegraphics[width=0.32\columnwidth]{../figs/fig_avg_distance_U0_V05_L150.pdf}
%    \includegraphics[width=0.32\columnwidth]{../figs/fig_avg_distance_U0_V05_L200.pdf}
%\end{figure}

\section{Trimer dynamics}
\label{sec:trimer_dynamics}

We study the propagation of a trimer by initially trapping the trimer of two heavy $\downarrow$-particles and one light $\uparrow$-particle in a small box of three lattice sites and then releasing it into a larger uniform lattice (quenching the box potential to zero). For this purpose, an additional site- and time-dependent potential term needs to be added to the Hamiltonian:
\begin{eqnarray}
H&=&-\sum_{\langle ij \rangle, \sigma}{t_{\sigma}(c_{i,\sigma}^{\dagger}c_{j,\sigma}+\mathrm{h.c.})}+\sum_in_{i, \uparrow}n_{i, \downarrow} \\ \nonumber &&+\sum_{\langle i,j \rangle, \sigma}V n_{i,\sigma}n_{j,\sigma}+\sum_{\langle i,j \rangle} V' n_{i, \uparrow}n_{j, \downarrow}+\sum_i{V_{\mathrm{trap}, i}(t)n_i}
\end{eqnarray}
where $V_{\mathrm{trap}, i}=0$ for $i$ corresponding to the three central lattice sites, and $V_{\mathrm{trap}, i}=50\,t_\uparrow$ for all other values of $i$ at $t=0$. 
After solving the ground state, the trap $V_{\mathrm{trap}, i}$ is switched off and the time evolution is obtained.

While the static analysis above was done for weakly interacting trimers, here we consider stronger interactions. 
The reasons for this are two-fold: trapping the atoms in a small box of three sites would have only a small overlap with the actual ground state wavefunction if the trimer size is very large, as shown in Fig.~\ref{fig:trimer_size}. Hence the trimer would be broken by the sudden release from the trapping potential. Secondly, as stronger interactions make the trimer smaller in size, we can more easily follow the propagation of the trimer by considering the movement of individual atoms.

\subsection{Trimer expansion}

\begin{figure}
  \includegraphics[width=0.49\columnwidth]{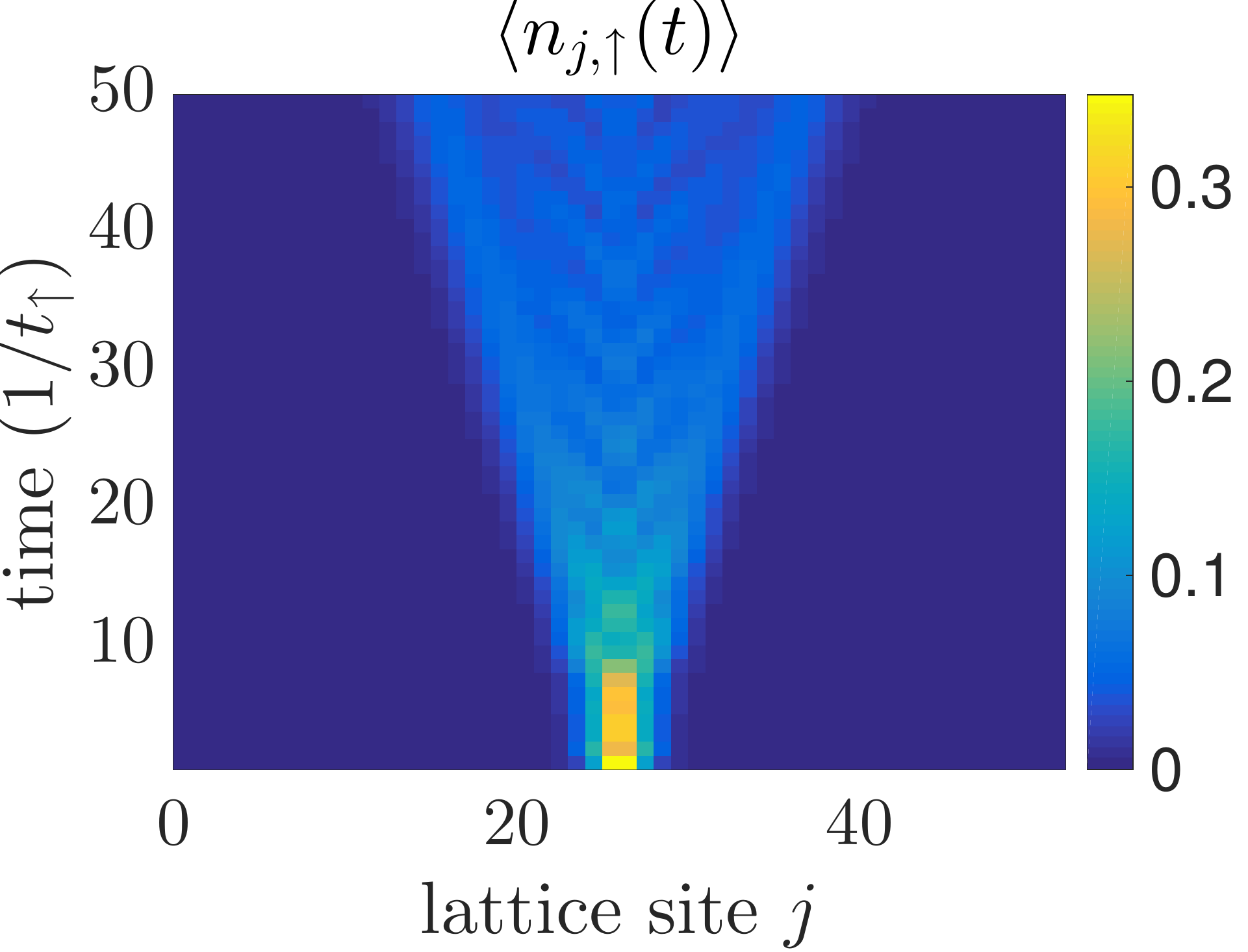}
  \includegraphics[width=0.49\columnwidth]{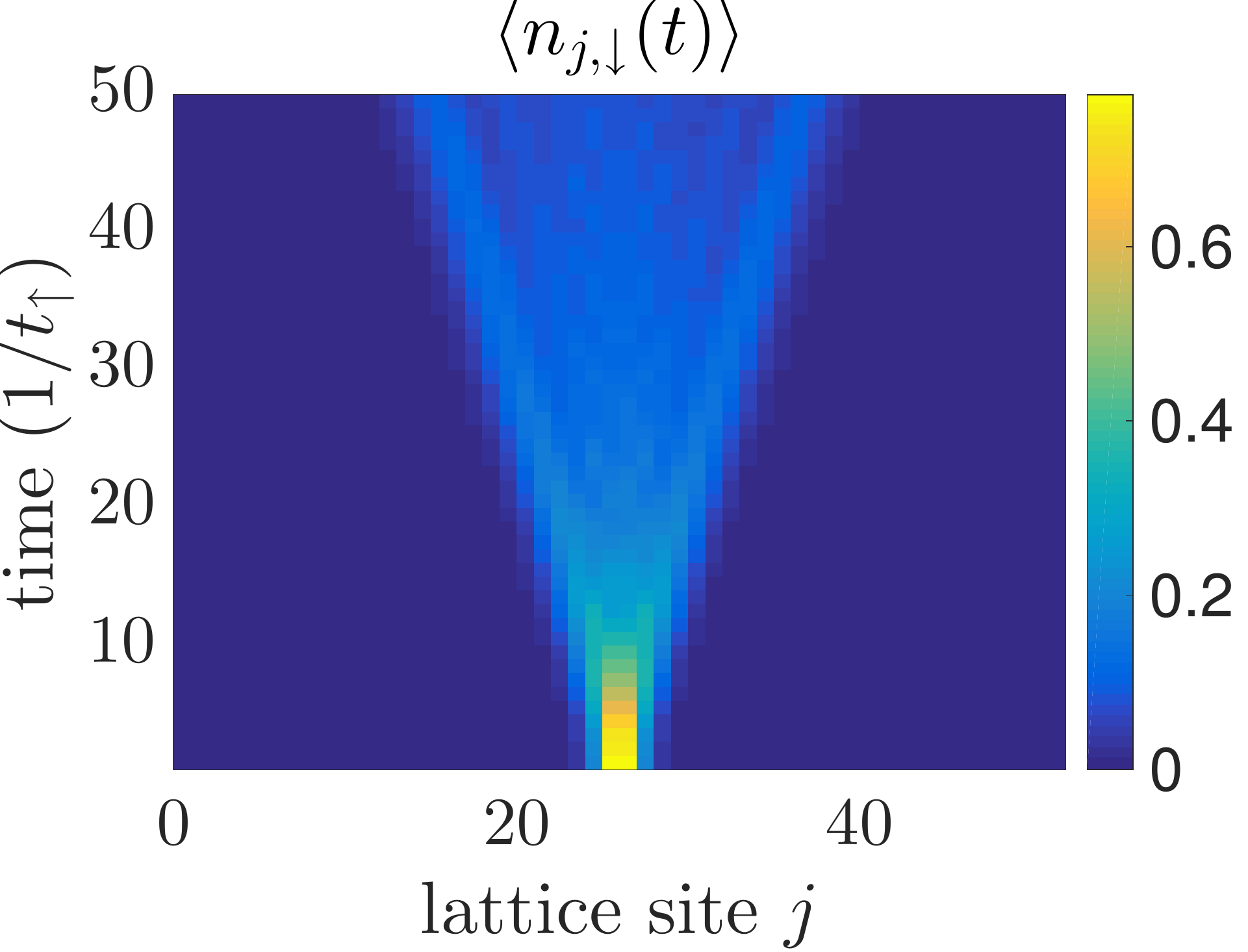}
  \includegraphics[width=0.49\columnwidth]{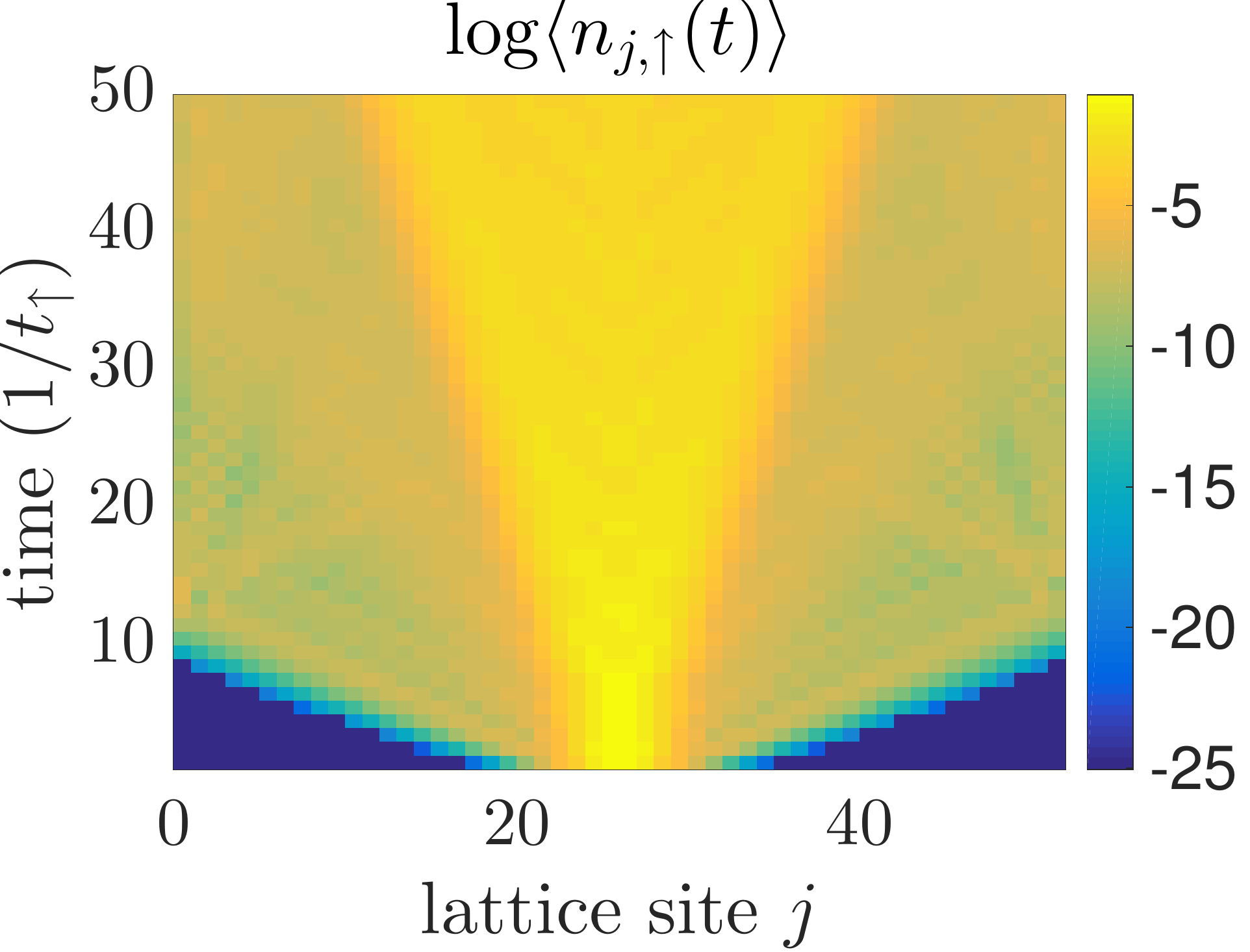}
  \includegraphics[width=0.49\columnwidth]{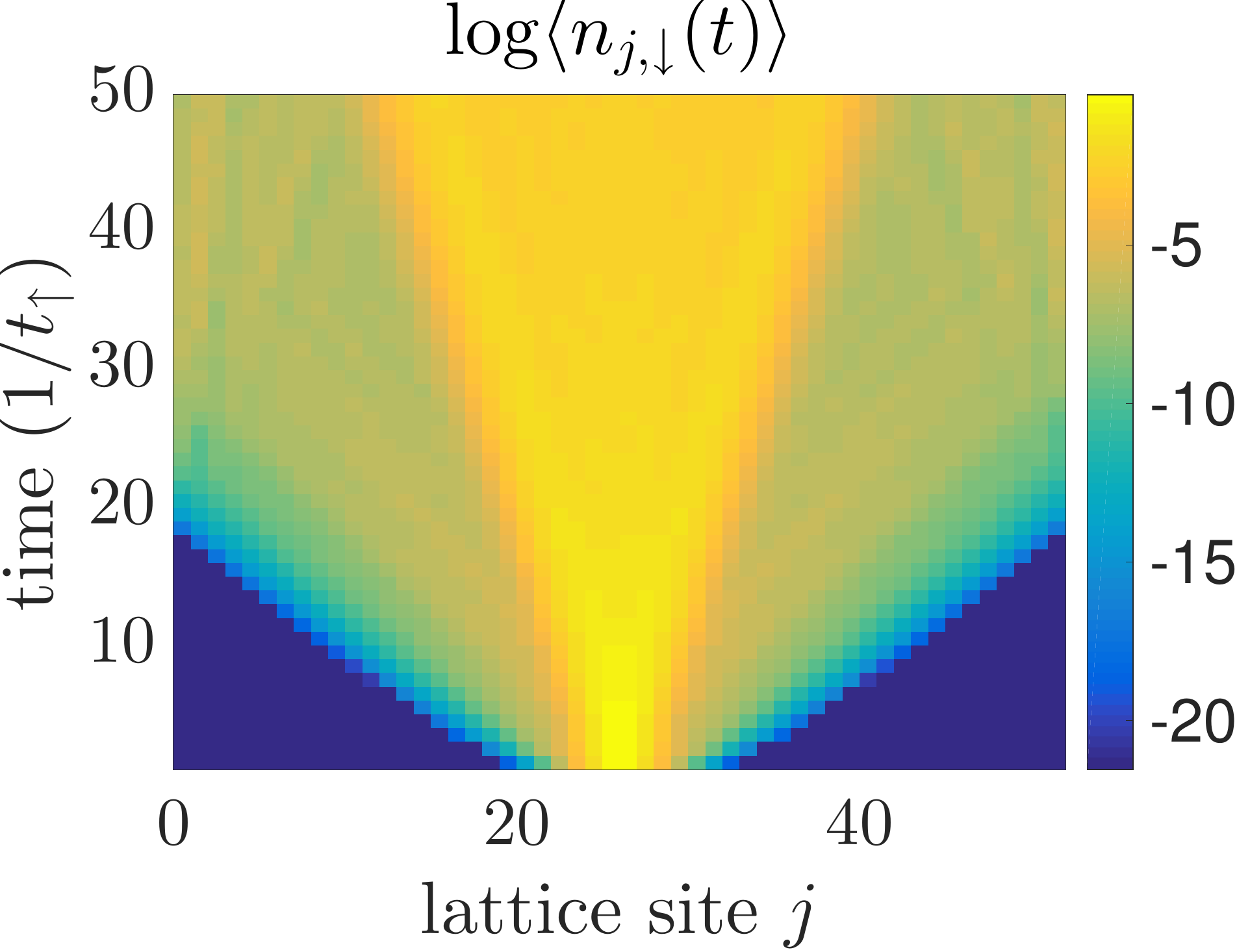}
  \caption{Expansion of atoms for $V=-2,U=0$. Top Left: density of the minority atom ($\uparrow$) as a function of time shows the expansion of the gas with a speed of roughly 13 sites in time 50 $\hbar/t_\uparrow$, i.e. 0.26 $t_\uparrow/\hbar$. Top Right: the density of the majority atoms ($\downarrow$) shows the same expansion.
    Bottom Left: The same minority atom density in logarithmic scale shows much smaller but also faster single-particle wave front that propagates at speed of roughly 20 sites in time 10 $\hbar/t_\uparrow$, i.e. speed of 2 $t_\uparrow/\hbar$. Bottom Right: The majority atom density in the logarithmic scale shows single-particle propagation speed of roughly 1 $t_\uparrow/\hbar$. Here hopping ratio is $t_\downarrow/t_\uparrow = 0.5$, which matches well with the observed single-particle propagation speeds.}
  \label{fig:wavefronts}
\end{figure}

Fig.~\ref{fig:wavefronts} shows the expansion of the initially trapped atoms out into the uniform lattice.
It shows three different propagation cones: two single-particle expansion cones corresponding to the hoppings of unpaired $\uparrow$ and $\downarrow$ atoms, and then much slower cone that shows in a similar way in both components.
The fast propagation cones are visible only in the logarithmic scale, but show clearly the expected single-particle propagation speeds of the two components. 
In the linear scale, only the slowly propagating cone is visible. Since this cone appears identically in both $\uparrow$ and $\downarrow$ densities, it suggests that the $\uparrow$ and $\downarrow$ atoms remain bound and the cone is due to the slow propagation of some kind of cluster.
However, a question remains: does the slow cone describe the propagation of a dimer or a trimer?

%{\color{red} What follows here should be redone: our picture of trimer propagation through doublon+singlon state is not correct since we have nearest-neighbour interactions also between same spin atoms. Either calculate again for $V'=0$ or do this again for $V=V'=-10$. The latter would be just calculating the average distance from the already calculated density matrices, right? And maybe do that also for $V=V'=-20$, and replace Fig.5 with those plots. Actually, we could maybe show the average distance for $V=-2$, $V=-10$ and $V=-20$.}

\begin{figure}
  \includegraphics[width=0.49\columnwidth]{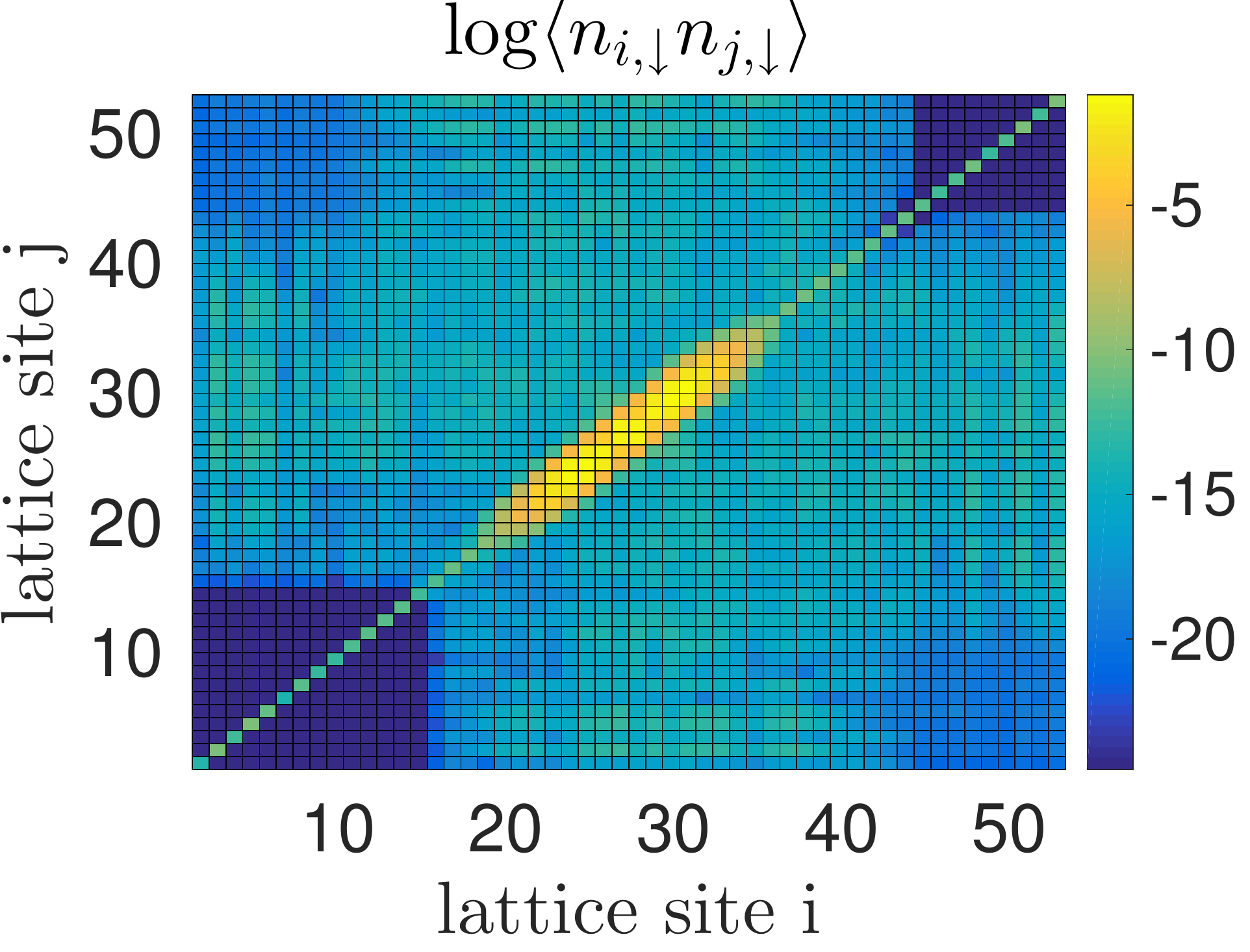}
  \includegraphics[width=0.49\columnwidth]{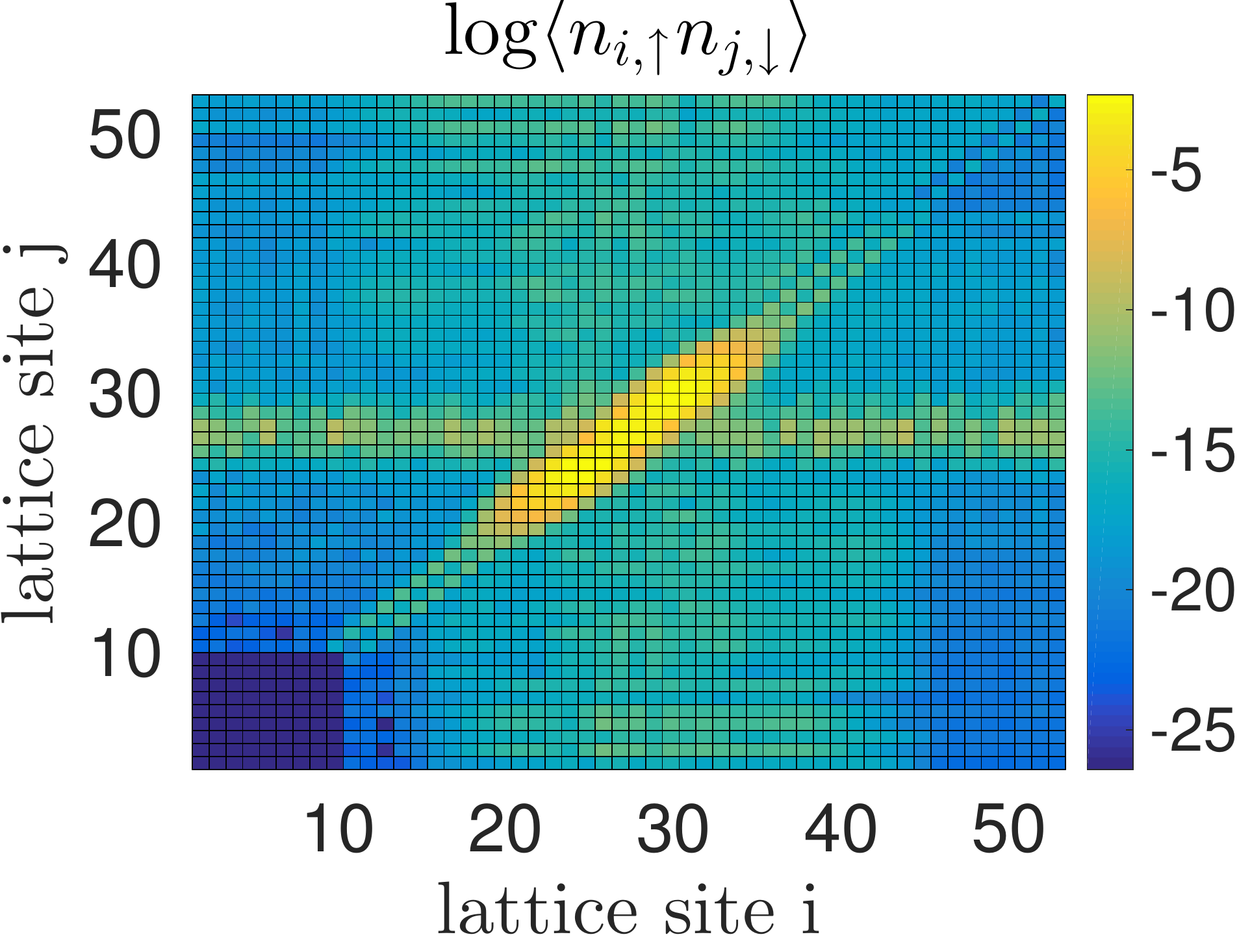}
  \caption{The $\log\langle n_{\downarrow,i} n_{\downarrow,j} \rangle$ (Left) and $\log\langle n_{\uparrow,i} n_{\downarrow,j} \rangle$ (Right) density matrices for $U=0.0, V=-10, t_{ratio}=0.3$ at time, $T=80$ show the broad diagonal contribution from trimer propagation and the much weaker dimer propagation appearing as a diagonal fork in the $\uparrow \downarrow$ correlator.}
  \label{fig:updncorrelator}
\end{figure}

To distinguish these two cluster configurations in the dynamics, higher order correlators need to be considered.
While actual three-body correlators are numerically very challenging to analyze, the combination of the two density-density correlators $\langle n_{i,\downarrow} n_{j,\downarrow}\rangle$ and $\langle n_{i,\uparrow} n_{j,\downarrow}\rangle$ will be sufficient. 

\subsection{Density-density correlators}

Fig.~\ref{fig:updncorrelator} shows the two density-density correlator matrices.
Both correlators show very prominent broad diagonal feature that has the same extent in both correlators.
Since the contributions near the diagonal describe nearby lying atoms, and since both $\downarrow-\downarrow$ and $\uparrow-\downarrow$ correlators show identical feature, it is very likely that this is due to trimers.
This interpretation is all the more convincing considering that the two dimers ($\downarrow-\downarrow$-dimer and $\uparrow-\downarrow$-dimer) have quite different propagation speeds due to different hopping rates of $\uparrow$- and $\downarrow$-atoms. 
Besides the broad trimer feature, one can also see weaker but faster (longer) diagonal contributions in both density matrices: a single diagonal line in $\downarrow-\downarrow$ correlator and a two-pronged diagonal fork in the $\uparrow-\downarrow$-correlator. The former describes simply tails of the single-particle density profile (which could arise either from $\downarrow$-singlons or $\uparrow-\downarrow$-dimers), and the latter is from nearest-neighbour $\uparrow-\downarrow$ dimers.
Here on-site interaction is vanishing $U=0$ but nearest-neighbour interaction is strong $V=-10\,t_\uparrow$, meaning that on-site dimers (doublons) are not stable.

%The density matrix shows also a clear ray along the horizontal direction. This ray corresponds to correlated state in which one $\downarrow$ atom has propagated far from the center of the lattice while the $\uparrow$ atom remains close to the initial place. 
%We associate this feature with broken trimer state in which the $\uparrow$ atom is still bound with a $\downarrow$ atom, and the other $\downarrow$ atom is ejected. 
%Since the $\uparrow$ atom remains close to the initial place, we can see that the $\downarrow \uparrow$-doublon propagation is very slow, whereas the lone $\downarrow$ atom propagates freely at the single-particle velocity.
%The $\langle n_{i,\uparrow} n_{j,\downarrow}\rangle$ density matrix shows thus clearly the difference between the $\downarrow \uparrow$-doublon propagation and the trimer propagation.
%{\color{red} It might be nice to connect this now with the $\downarrow \downarrow$ density matrix, as it should show the $\downarrow \downarrow$ doublon as a narrow ray and the same kind of diagonal as here from the trimer.}

From the density matrices, such as the ones shown in Fig.~\ref{fig:updncorrelator}, one can determine the propagation speed of trimers, but also to identify the actual process through which the trimer propagates. 
The diagonal elements in the density matrix $\langle n_{i,\uparrow} n_{i,\downarrow}\rangle$ correspond to the on-site doublon density distribution in the lattice.
Since $\downarrow \uparrow$-dimers are bound only through the nearest-neighbour interaction (since on-site interaction $U=0$), they do not contribute to the on-site doublon density. 
In contrast, trimers provide the main contribution for that correlator, and hence we can use it for quantifying the trimer propagation speed. 

\subsection{Trimer propagation speed}

\begin{figure}
  \includegraphics[width=0.49\columnwidth]{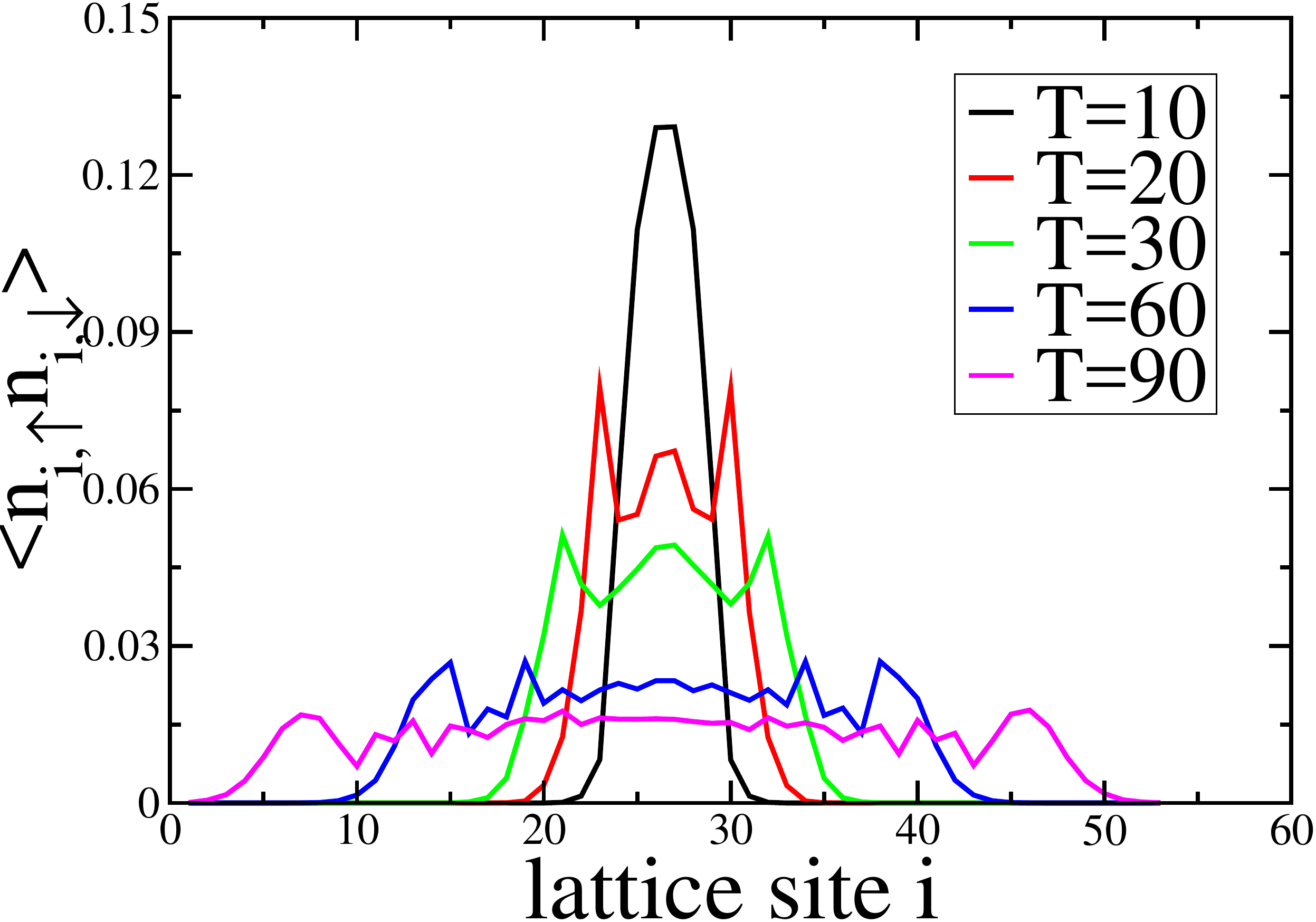}
  \includegraphics[width=0.49\columnwidth]{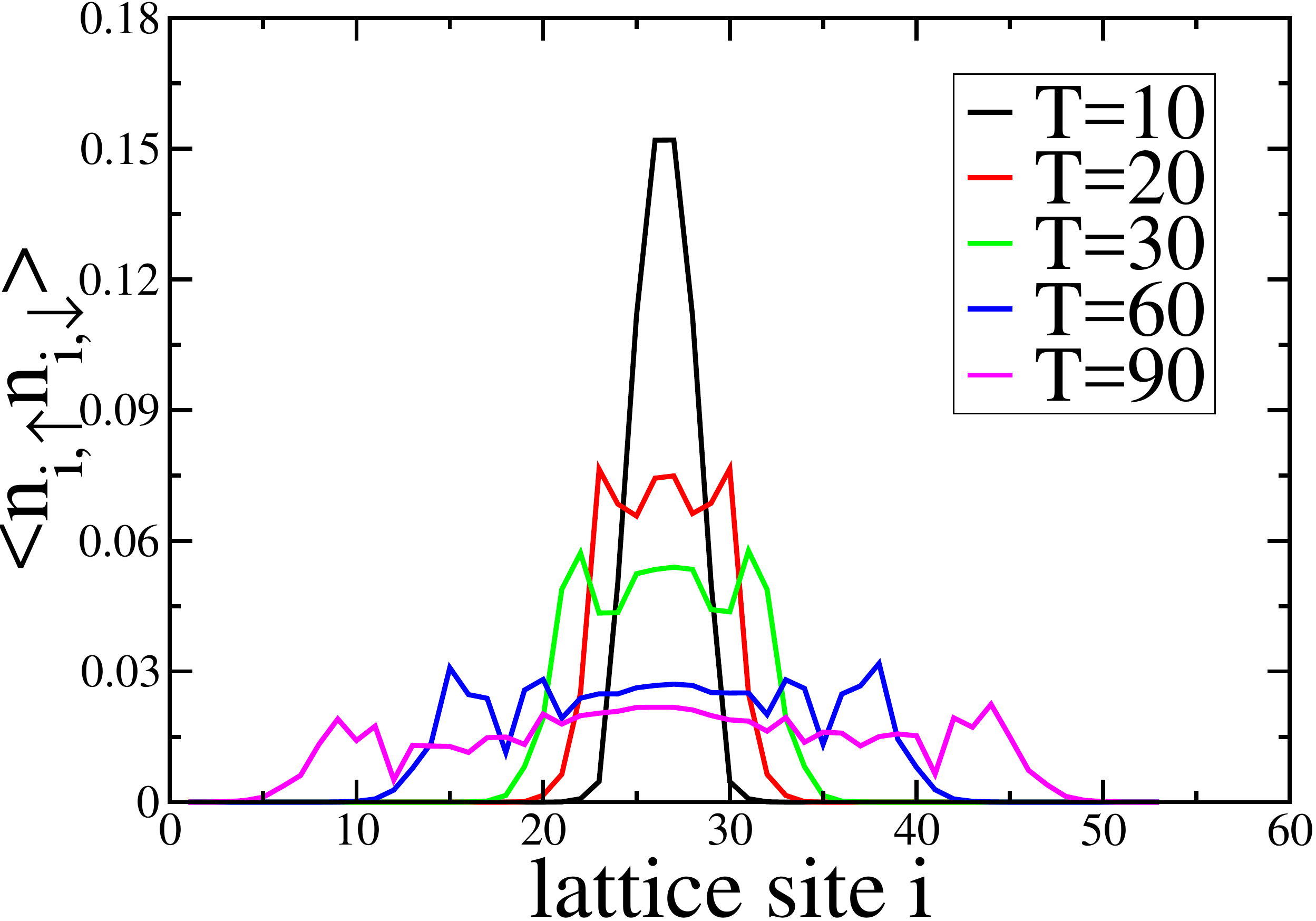}
  \caption{Trimer dynamics: the diagonal elements of the $\uparrow$-$\downarrow$ correlator, $\langle n_{\uparrow,i} n_{\downarrow,i} \rangle$ for $U=0.0, t_{ratio}=0.7, V=-10.0$ (left) and $V=-20.0$ (right) at different times after the release from the initial strong confinement, shows the trimer propagation wavefront.}
  \label{fig:updndistance}
\end{figure}
\begin{figure}
  \includegraphics[width=0.9\columnwidth]{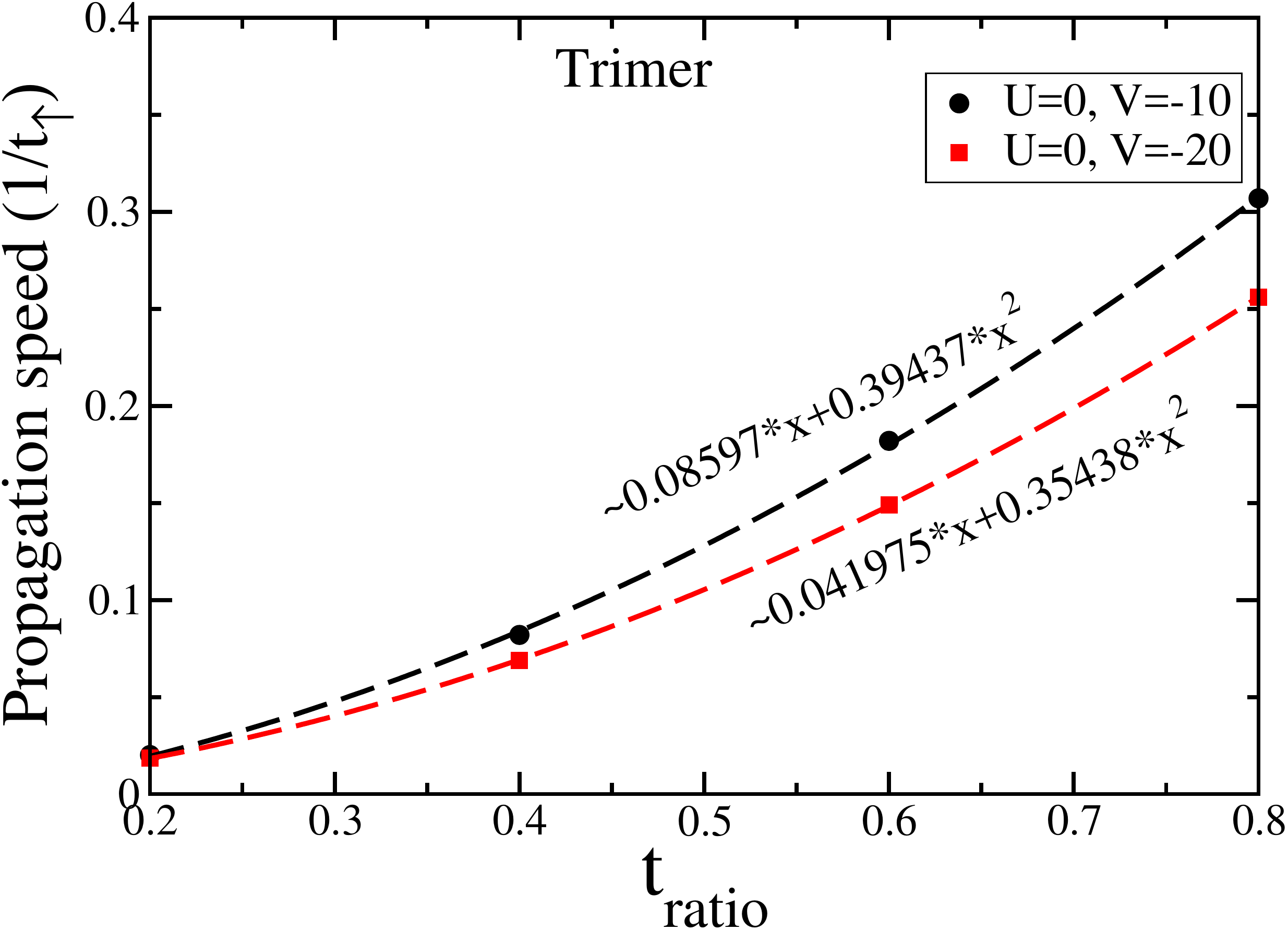}
  \caption{Trimer propagation speed as a function of the hopping ratio $t_\downarrow/t_\uparrow$ for nearest-neighbour interaction strengths $V=-10$ and $V=-20$. Quadratic fits show the clear quadratic dependence on the hopping ratio, but only a weak dependence on the interaction strength $V$.} 
  \label{fig:trimer_speed}
\end{figure}

Fig.~\ref{fig:updndistance} shows the diagonal elements of the $\uparrow$-$\downarrow$-correlator, i.e. the on-site doublon density distribution.
%{\color{red}\ref{fig:null1}WE SHOULD ALSO PLOT THE UPARROW-DENSITY PROFILES AND THE DOWNARROW DENSITY PROFILES FOR THE SAME DATA SETS. WE SHOULD SEE THE SAME EXTENT IN THOSE AS WELL.}
%\begin{figure}
%  \includegraphics[width=0.9\columnwidth]{fig_null.pdf}
%  \caption{THIS SHOULD BE UPARROW-DENSITY PROFILES AND DOWNARROW-DENSITY PROFILES FOR THE SAME DATA SETS.} 
%  \label{fig:null1}
%\end{figure}
From these plots, we can determine the distance that the trimer has propagated by calculating the full-width at half-maximum (FWHM) for a given time step.
Although the profiles here are far from Gaussian distributions, the FWHM does provide a simple and transparent measure of the extent of the distribution.
For strongly bound trimers, the propagation speed is found to be only weakly dependent on the nearest-neighbour interaction $V$ (for $V=V'$) and scales roughly as a square of the hopping ratio $t_\mathrm{ratio}^2$, see Fig.~\ref{fig:trimer_speed}.

\begin{figure}
  \includegraphics[width=0.9\columnwidth]{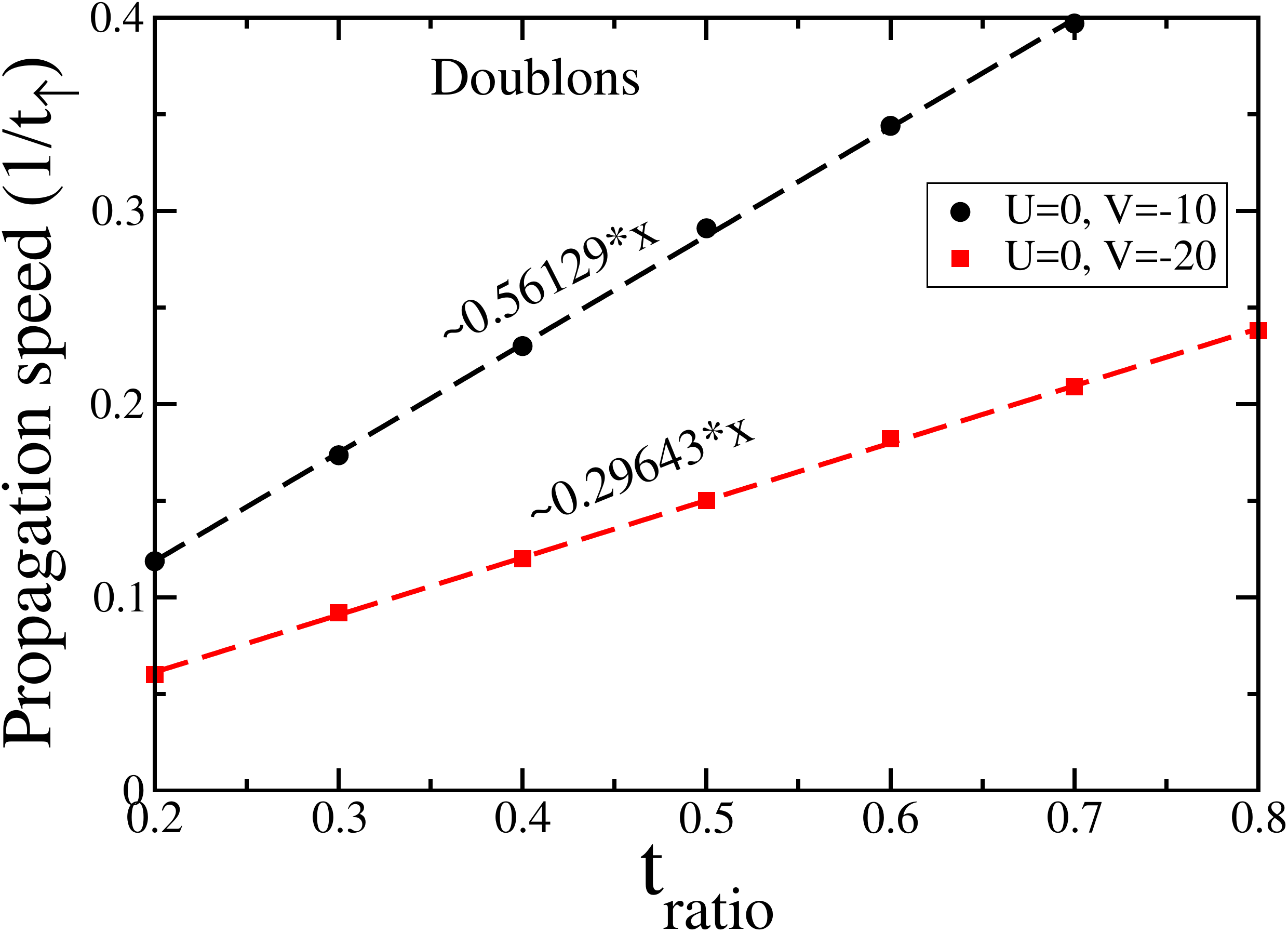}
  \caption{Doublon propagation speed as a function of the hopping ratio $t_\downarrow/t_\uparrow$ for nearest-neighbour interaction strengths $V=-10$ and $V=-20$. The doublon speed depends linearly on the hopping ratio, but it also depends strongly on the interaction strength. For strong interactions, the speed is inversely proportional to $V$.} 
  \label{fig:doublon_speed}
\end{figure}
This is quite different from the propagation of dimers, whose propagation speed scales as $\sim \frac{1}{V}$, see Fig.~\ref{fig:doublon_speed}.
Notice that the slope of the fitted linear function (very close to 0.3 for $V=-20\,t_\uparrow$) differs from what would be expected for on-site dimers.
For localized dimers interacting through strong on-site interaction $U$, the propagation speed is $4t_\uparrow t_\downarrow/U$, where the denominator $U$ comes from the energy of the virtual intermediate state and the prefactor $4$ arises from two different orderings in which the transport of the dimer can take place.
In the case of nearest-neighbour interactions (with $U=0$), one additional process allows the dimer to move: a nearest-neighbour dimer having $\uparrow$-atom in the left site and $\downarrow$-atom in the right site can hop to the right by having the $\uparrow$-atom hop twice to the right. This changes the orientation of the dimer from $\uparrow \downarrow$-configuration to $\downarrow \uparrow$-configuration, effectively utilizing the degeneracy of the dimer state. Having three possible orderings for the transport of the dimer instead of two yields propagation speed of $6 t_\uparrow t_\downarrow/V$. For $V=-20$, this yields slope $6/20=0.3$ as seen in Fig.~\ref{fig:doublon_speed}.

We have been unable to formulate similar simple description for the trimer propagation. As will be seen below, trimer propagation occurs through near-degenerate states, and it does not utilize intermediate virtual states with high energy cost.
Indeed, for sufficiently strong interactions, trimers propagate faster than dimers, as seen by comparing Figs.~\ref{fig:trimer_speed} and~\ref{fig:doublon_speed}.

\begin{figure}
  \includegraphics[width=0.47\columnwidth]{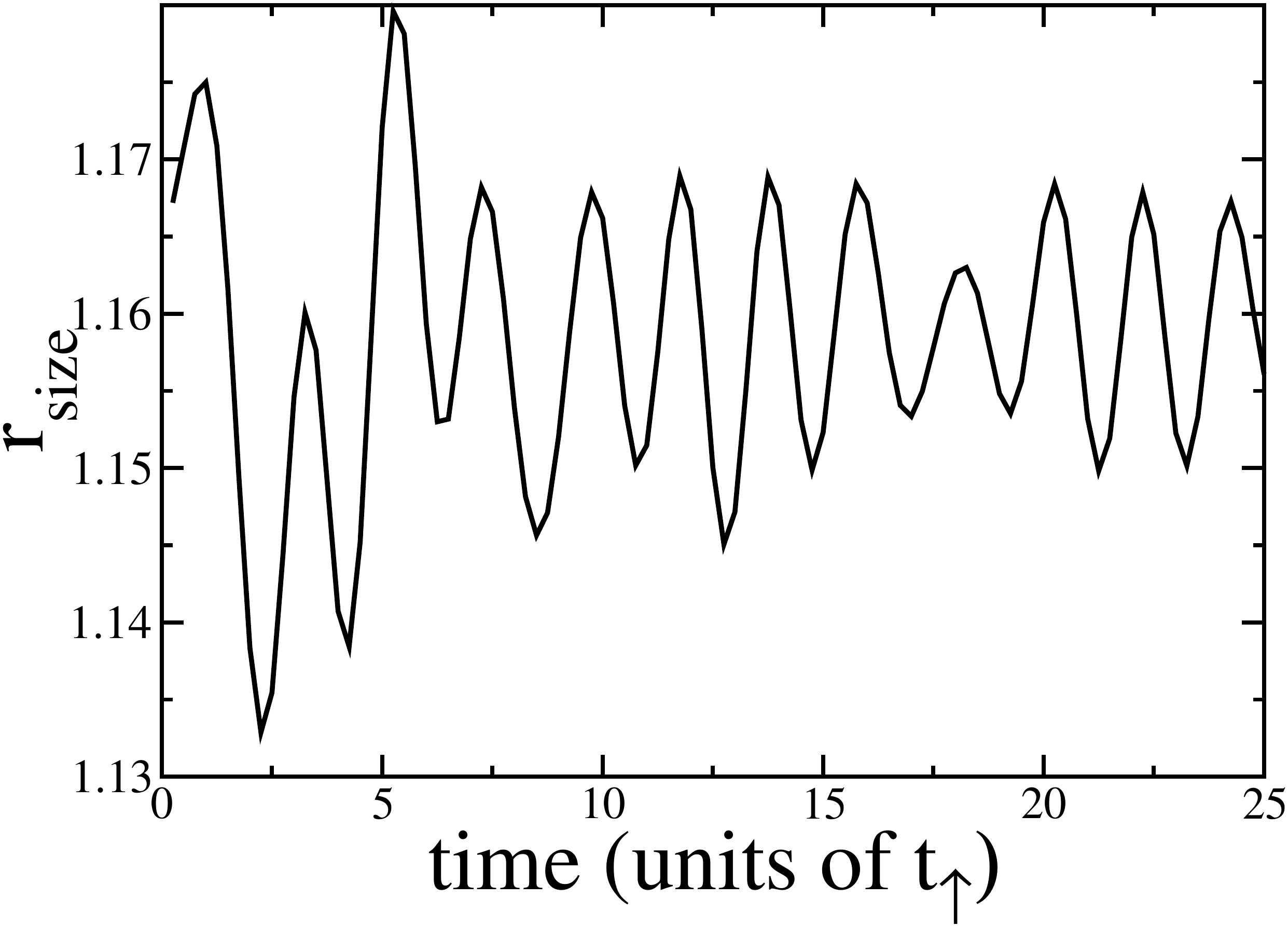}
  \includegraphics[width=0.47\columnwidth]{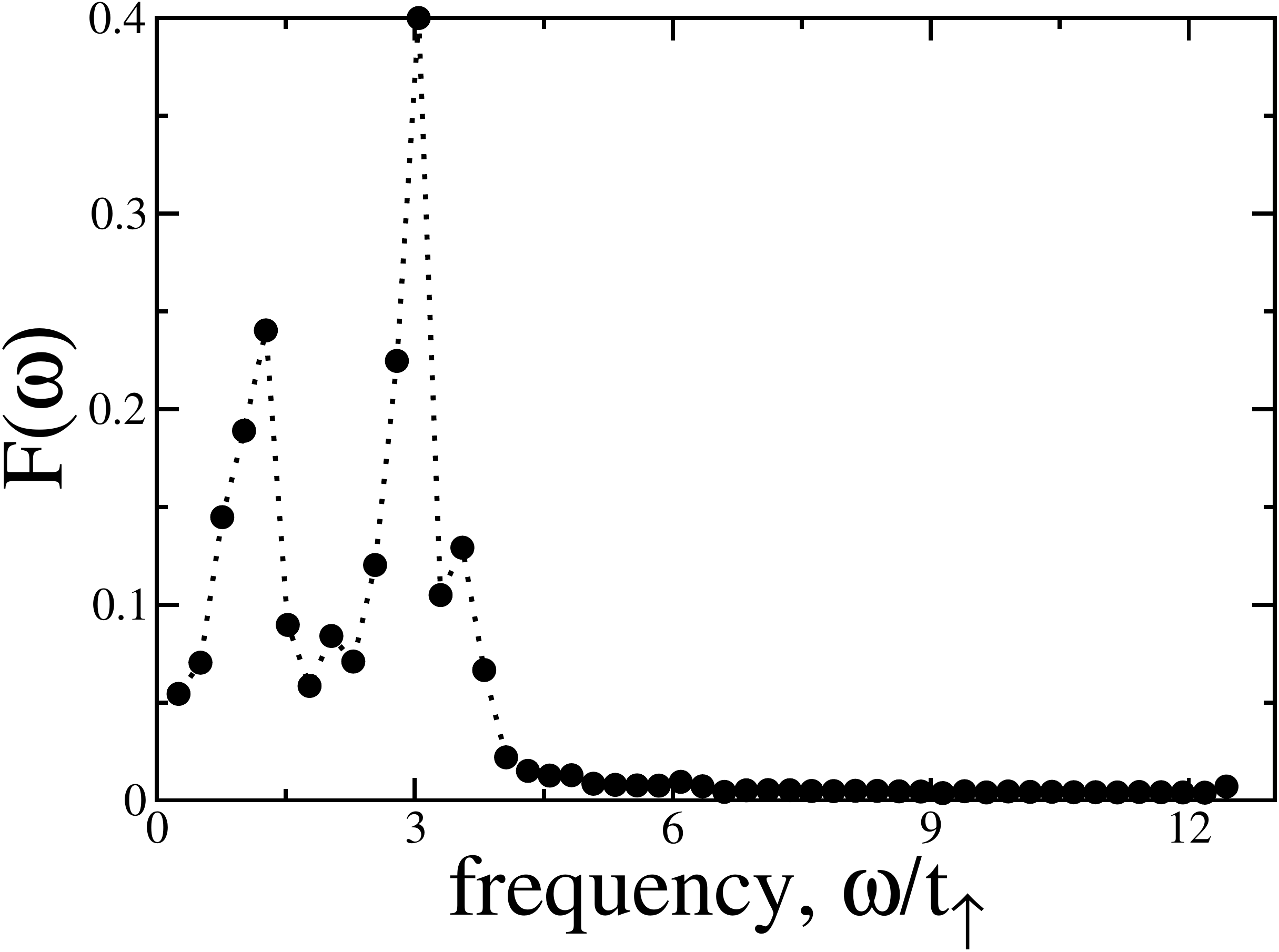}
  %downfig_dndndistance_oscillations.pdf}
%  \includegraphics[width=0.95\columnwidth]{plot-avge-distance-down-down-vs-time-U0_V-20_Vprime-20_tratio0_7_L53-Lbox3-finer-resolution.png}%{fig_dndndistance_oscillations_tr07.pdf}
  \caption{Left: the average distance between the two $\downarrow$-atoms shows oscillations corresponding to transitions between different trimer states for $U=0.0, V=-20.0$ and $t_{ratio}=0.7$. Right: Fourier transform of the average distance shows strong peaks at frequencies $\omega = 3.0\,t_\uparrow$ and $\omega=1.3\,t_\uparrow$, yielding energy separations of the various trimer states.}
  \label{fig:dndndistance}
\end{figure}

\subsection{Model for trimer propagation}

The surprising faster propagation speed (for sufficiently strong interactions) of trimers can be
analyzed further by considering the various two-particle correlators.
Fig.~\ref{fig:dndndistance} shows the time-dependence of the average distance between the two $\downarrow$-atoms, i.e. the average distance in Eq.~\eqref{eq:trimer_size} calculated for subsequent time steps.
Figure shows first of all that the average distance is and remains quite small, showing that trimers are very strongly bound and they survive the release from the box with very high probability.
It also shows clear oscillations that, with the help of Fourier transform, can be seen to correspond to a few dominant frequencies $\omega \approx 3.05\,t_\uparrow$ and $1.27\,t_\uparrow$.
These frequencies correspond to energy separations of transitions between various trimer states.
\begin{figure}
\centering
  \includegraphics[width=0.95\columnwidth]{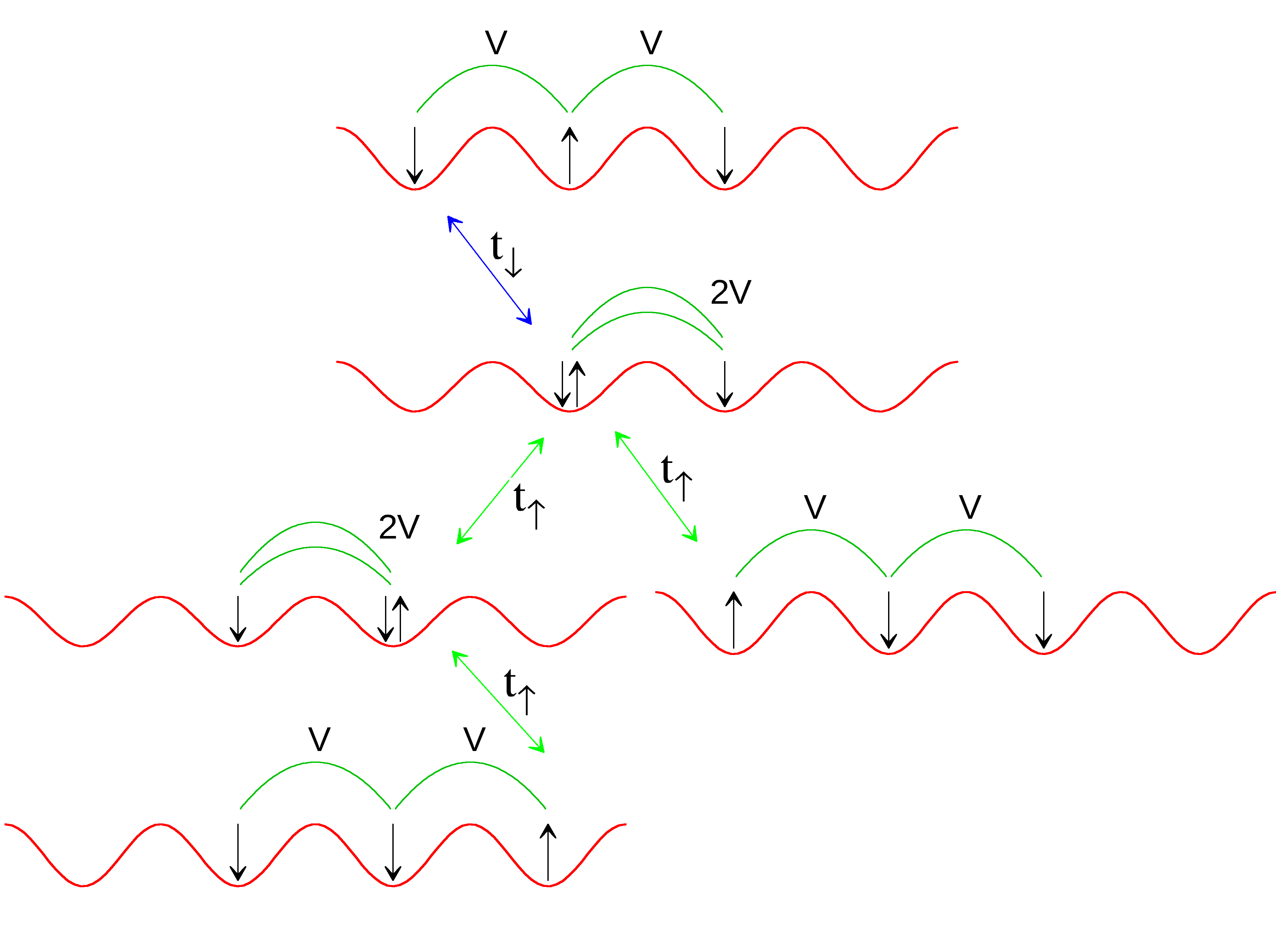}
  \caption{Degenerate states of deeply bound trimers for equal nearest-neighbour interactions between like and opposite spins ($V=V'$) and in absence of on-site interactions ($U=0$). All configurations have the total interaction energy of $2V$. The topmost state is connected to the next state through hopping of the $\downarrow$-atom, but the rest of the states are connected through the hopping of the $\uparrow$-atom. In addition to the states shown here, the mirror images of the four lower configurations are also degenerate in a symmetric way, and translationally symmetric states have been neglected (i.e. states with the same configuration as the ones shown here, but shifted along the lattice.)}
  \label{fig:schematic_trimer_states}
\end{figure}

Fig.~\ref{fig:schematic_trimer_states} shows five different trimer configurations that all have the same interaction energy of $2V$ in absence of on-site interaction $U$. The states are coupled through hopping of the $\uparrow$-atom or either of the $\downarrow$-atoms. 
These couplings hybridize the states, but the eigenstates and -energies can be easily obtained within this deeply bound limit. 
It is the separations between these eigenenergies that is observed in the oscillation frequencies of Fig.~\ref{fig:dndndistance}.
The model can be made even simpler by observing that the correlator for the distance between the two $\downarrow$-atoms is sensitive only to the hoppings of the $\downarrow$-atoms, i.e. to the transitions of between the two topmost states in Fig.~\ref{fig:schematic_trimer_states}.
We can thus artificially separate the space spanned by the trimer configurations depicted in Fig.~\ref{fig:schematic_trimer_states} in two subspaces corresponding to different $\downarrow$-atom configurations. One of the subspaces consists only of the topmost state (the symmetric $\downarrow \uparrow \downarrow$-state) in Fig.~\ref{fig:schematic_trimer_states}, whereas the other subspace is spanned by all other states, which are all coupled by the hopping of the $\uparrow$-atom. Solving the eigenspectrum of the latter yields eigenenergies $\pm 0.618\,t_\uparrow$ and $\pm 1.618\,t_\uparrow$ and the corresponding eigenstates $\Psi_n$. These eigenstates $\Psi_n$ are then coupled with the symmetric $\downarrow \uparrow \downarrow$-state through the coupling $t_\downarrow$. At the simplest level these couplings can be described as a collection of two-level systems coupling the symmetric $\downarrow \uparrow \downarrow$-state with each of the eigenstates $\Psi_n$ separately. That is, the couplings are described by the Hamiltonian
\begin{equation}
   \begin{pmatrix}
     0 & t_\downarrow\\
     t_\downarrow & E_n
   \end{pmatrix},
\end{equation}
where $E_n$ is the eigenenergy of the eigenstate $\Psi_n$. For $t_\downarrow = 0.7\,t_\uparrow$, we finally obtain eight eigenenergies for the full system: $\pm 0.2608\,t_\uparrow$, $\pm 0.4562\,t_\uparrow$, $\pm 1.074\,t_\uparrow$, and $\pm 1.879\,t_\uparrow$.
While not all of the energy separation combinations can be resolved in the Fourier spectrum in Fig.~\ref{fig:dndndistance}, the dominant peaks can be easily identified as the separations of $1.879\,t_\uparrow - (-1.074\,t_\uparrow) \approx 2.94\,t_\uparrow$, and $0.2608\,t_\uparrow - (-1.074\,t_\uparrow) \approx 1.33\,t_\uparrow$.

\begin{figure}
  \includegraphics[width=0.35\columnwidth]{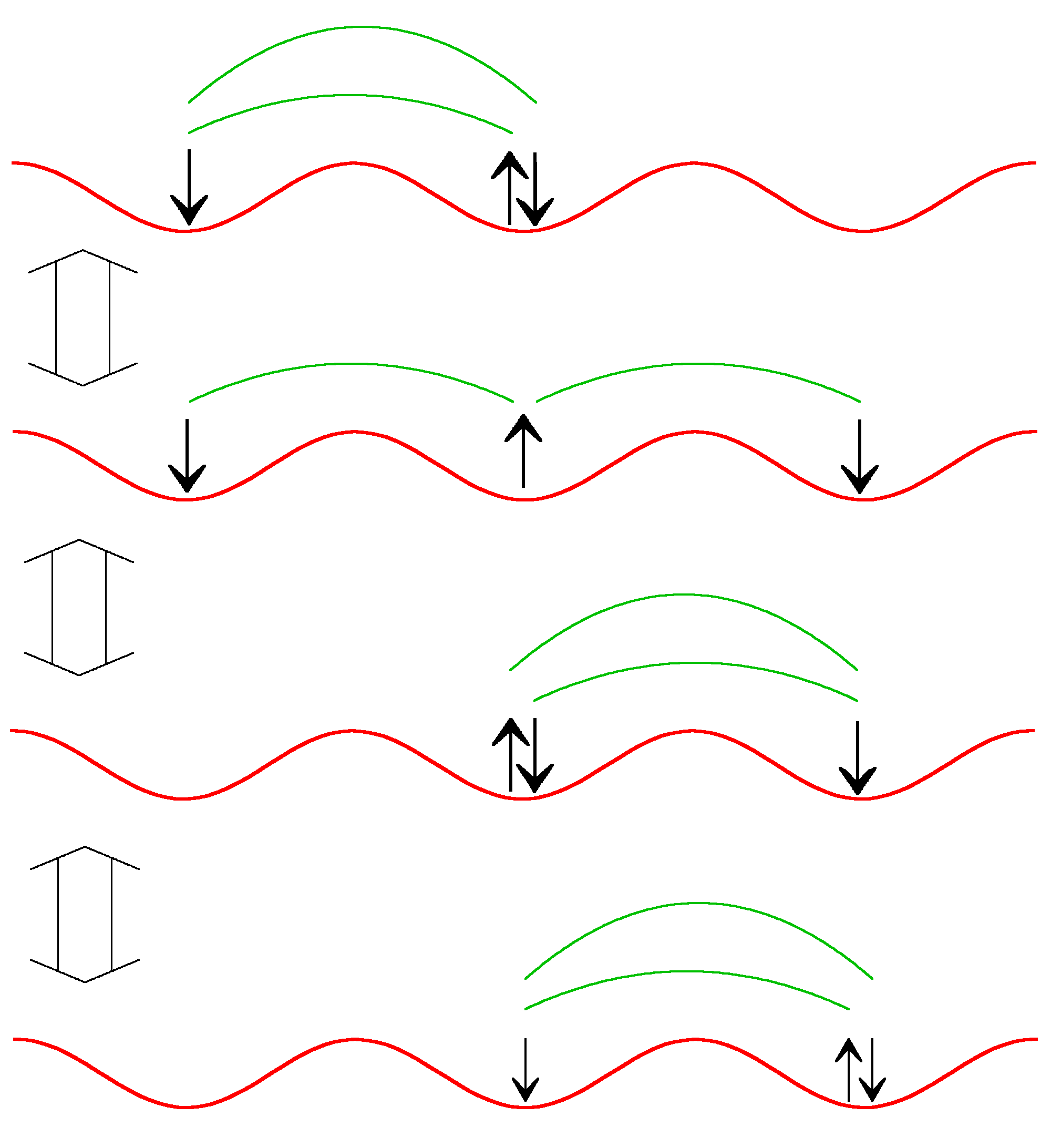}
  \includegraphics[width=0.63\columnwidth]{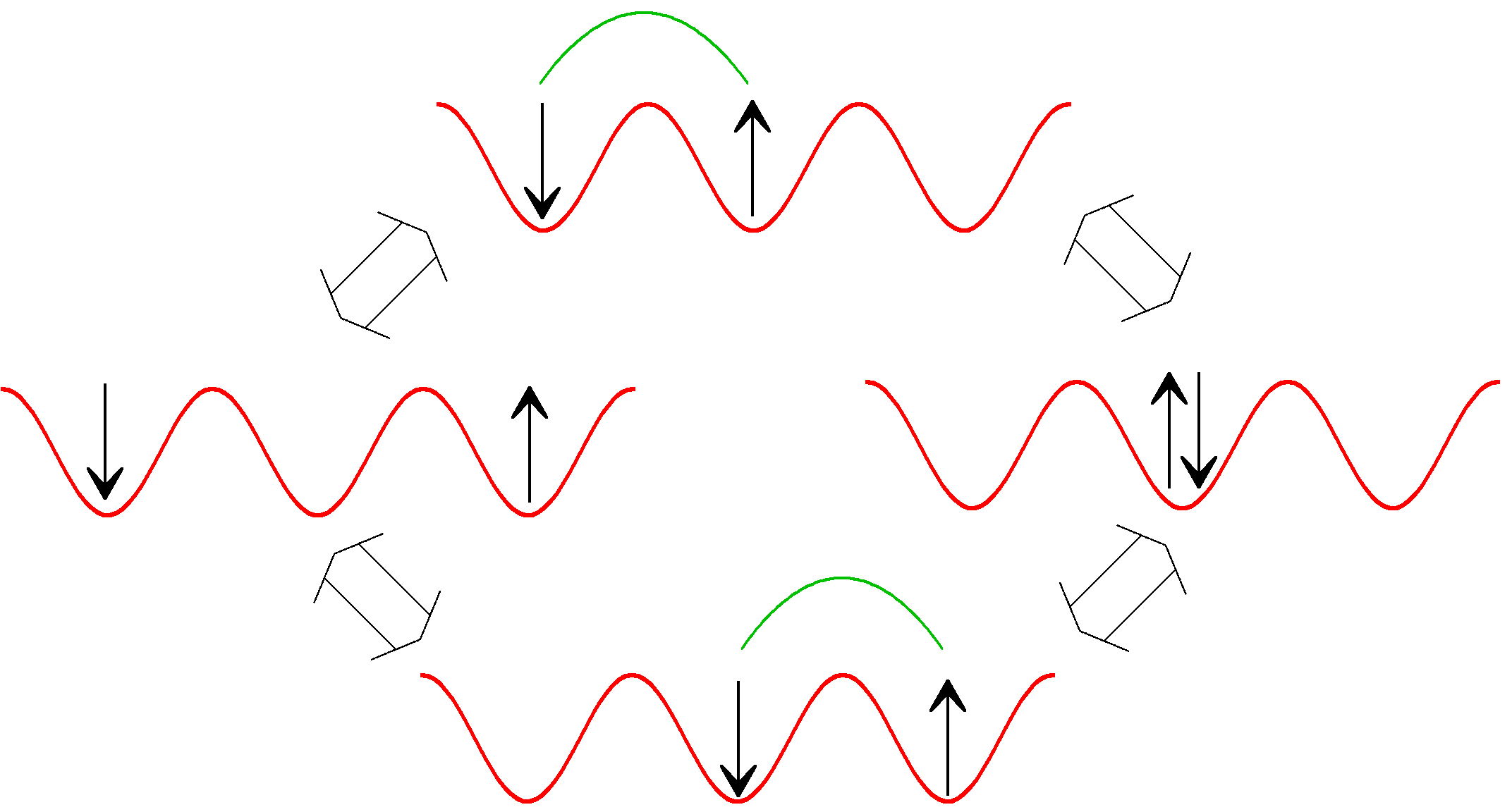}  
  \caption{Left: The near-degeneracy of the trimer configurations shown in Fig.~\ref{fig:schematic_trimer_states} allows the trimers to propagate without intermediate virtual states involving high energy cost. The starting configuration is the same as the final configuration, except for a single lattice site shift to the right along the lattice.
  Right: For sufficiently large interactions $V \gg t_\uparrow$, the propagation of dimers is slower than for trimers, since dimer propagation necessarily involves virtual states with a broken nearest-neighbour dimer.} 
  \label{fig:schematic_trimer_propagation}
\end{figure}

The simple explanation for the rapid propagation of the trimers in the strongly interacting limit is thus that the trimer propagation can occur through various near-degenerate trimer configurations, as shown in Fig.~\ref{fig:schematic_trimer_propagation}. In contrast, dimer necessarily has to propagate via virtual states involving broken pairs either through on-site doublon state or by next-nearest-neighbour configuration. 
This analysis can be extended to larger clusters as well. 
%Interestingly, for two-component fermions and with large nearest-neighbour interaction strength $V$, trimers appear to be the only multiparticle clusters that allow such rapid propagation in one-dimension.

\section{Experimental considerations}
\label{sec:experiments}

\subsection{Nearest-neighbor interactions and dipolar atoms}

The choice of equal intra- and interspecies nearest-neighbor interactions $V=V'$ probably describe best systems in which the two fermionic components are identical atoms in different hyperfine states, such as erbium $^{167}$Er- or $^{161}$Dy-atoms, both having strong magnetic moments of $7\mu_B$ and $10\mu_B$ respectively, and hence potential for strong nearest-neighbor interactions. 
The mass imbalance, which is not crucial for many of the phenomena considered in this work, can be realized by state-dependent optical lattices~\cite{Mandel_coherent_2003} or by utilizing magnetic modulation~\cite{Jotzu_creating_2015}.
Strong nearest-neighbor interactions pose also further complications, as the next-nearest neighbor interactions will become important. 
This will provide a still broader spectrum of bound states and will enhance the formation of larger clusters. 
In such case, the rather short-sized trimers considered here in the dynamic studies, will provide only a subset of interesting multimer dynamics.

As strong attractive long-range interactions tend to favor large cluster formation, reaching low temperatures in the experiment is not required. 
Indeed, even an equilibrium state may be unwanted, and the interesting trimer propagation physics considered here may be most easily studied by performing an interaction quench from a weakly interacting regime to strong interactions.

\subsection{Role of on-site interactions}

The analysis here was done mainly for the case of vanishing on-site interactions $U=0$.
While the on-site interaction can be tuned using Feshbach resonances~\cite{Stan_observation_2004,Inouye_observation_2004,Ferlaino_feshbach_2006,Ferlaino_erratum_2006,Wille_exploring_2008,Voigt_ultracold_2009}, complete suppression of the interaction is unlikely in actual experiments, especially if the nearest-neighbor interaction is strong. 
Adding on-site interactions will lift the degeneracy of the five trimer configurations in Fig.~\ref{fig:schematic_trimer_states}. Such addition can easily be incorporated in the simple model.
However, as seen in the above model, the degeneracy of the trimer spectrum is already lifted by the particle hoppings,
and thus no qualitative changes is expected as long as the on-site interaction $U$ is much weaker than the nearest-neighbour interaction $V$.

\subsection{Detecting trimers}

This work considered various one- and two-particle observables for analyzing trimer stability and propagation.
It is likely that experimental works on trimer formation would first involve strongly bound trimers, and hence large and observable trimer gap. This can be probed by various spectroscopic methods~\cite{tarruell}. However, the average distance between the $\downarrow$-atoms in a dilute sample should be observable in ultracold gas microscope setups~\cite{edge_imaging_2015,omran_microscopic_2015,cheuk_quantum-gas_2015,haller_single-atom_2015,parsons_site-resolved_2015,Mitra2017}, and thus provide a way to detect weakly bound trimers.

Even if the true many-body ground state would involve large atom clusters, in practice the trimers and other smaller clusters will be present as excited states.
In such a case the trimers should be clearly observable, due to their high propagation speed. 
This is particularly the case when the trimers (and singlons) are the only rapidly propagating entities, as larger clusters and dimers are slowed down by the required virtual intermediate states involving high energy costs.

\section{Conclusion}
\label{sec:conclusion}

In conclusion, we have studied the two-component one-dimensional extended Fermi-Hubbard model with nearest-neighbor interactions. In addition, we included mass imbalance between the two fermionic species. We focused on the static and dynamic properties of trimers. By looking at the energy spectrum, we redefined the measure of trimer gap generalizing it to systems with nearest-neighbor interactions. 
The spectroscopic analysis of the trimer state was supplemented by the study of two-particle correlators, providing a measure of the size of the trimer. The two methods provide a way for experimental observation of trimers.

To understand how trimers propagate, we looked at relevant one- and two-point correlators. We observed different wavefronts propagating at different speeds, and attributed them to the single particle, doublon and trimer motion. We further investigated the propagation speeds, and concluded that the trimers move faster than the doublons in the strong interaction limit. 
We provided a model to explain this behavior, underlying the role played by the near degeneracy of the trimer ground state. 
Finally we discussed experimental realization with ultracold dipolar atoms in optical lattices by which our predictions can be verified.

\acknowledgements{A.D. would like to thank Daniel Jaschke for discussions on using OSMPS code. This work was supported by the Academy of
Finland through its Centers of Excellence Programme (2012-2017) and under Project Nos. 284621, 303351 and 307419, and by the European Research Council (ERC-2013-AdG-340748-CODE). Computing resources were provided by CSC - the Finnish IT Centre for Science.

\bibliographystyle{unsrt}
\bibliography{references} 

\end{document}